\documentclass[11pt,aps,pre,notitlepage,tightenlines,longbibliography]{revtex4}
\usepackage{graphicx}
\usepackage{rotating}
\usepackage{amsthm}
\usepackage{amsmath}
\usepackage{amssymb}
\usepackage{bm}
\usepackage{subfigure}
\usepackage[toc,page,title,titletoc,header]{appendix}



\newcommand{\beq}{\begin{equation}}
\newcommand{\eeq}{\end{equation}}

\def\A{{\bf A}}

\def\D{{\bf D}}
\def\E{{\bf E}}

\def\F{{\bf F}}
\def\g{{\bf g}}
\def\k{{\bf k}}
\def\I{{\bf I}}
\def\L{{\bf L}}
\def\m{{\bf m}}
\def\n{{\bf n}}
\def\p{{\bf p}}
\def\q{{\bf q}}
\def\Q{{\bf Q}}
\def\r{{\bf r}}
\def\R{{\bf R}}
\def\S{{\bf S}}
\def\T{{\bf T}}
\def\u{{\bf u}}
\def\v{{\bf v}}
\def\x{{\bf x}}
\def\X{{\bf X}}
\def\I{{\bf I}}
\def\z{{\bf z}}
\def\0{{\bf 0}}
\def\curl{{\nabla \times}}

\def\dotp{\dot{\bf p}}
\def\gradu{\nabla \u}

\def\barphi{{\bar\phi}}
\def\balp{{\bm \alpha}}
\def\bsig{{\bm \sigma}}

\def\bomega{{\bm \omega}}
\def\bPhi{{\bm \Phi}}

\def\zhat{{\bf{\hat z}}}
\def\khat{{\bf{\hat k}}}
\def\[{\left [}
\def\{{\left {}
\def\}{\right }}
\def\]{\right ]}

\begin{document}
\title{Analytical structure, dynamics, and coarse-graining of a
kinetic model of an active fluid}
\author{Tong Gao}
\affiliation{Department of Mechanical Engineering \\
Department of Computational Mathematics, Science and Engineering, Michigan State University}
\author{Meredith D. Betterton}
\affiliation{Department of Physics, University of Colorado at Boulder}
\author{An-Sheng Jhang}
\affiliation{Courant Institute of Mathematical Sciences,
  New York University}
\author{Michael J. Shelley}
\affiliation{Courant Institute of Mathematical Sciences,
  New York University \\
  Center for Computational Biology, Flatiron Institute, New York}
\date{\today}
\begin{abstract}
  We analyze one of the simplest active suspensions with complex
  dynamics: a suspension of immotile ``Extensor'' particles that exert
  active extensile dipolar stresses on the fluid in which they are immersed.
  This is relevant to several experimental systems, such as recently
  studied tripartite rods that create extensile flows by consuming a
  chemical fuel. We first describe the system through a Doi-Onsager
  kinetic theory based on microscopic modeling. This theory captures
  the active stresses produced by the particles that can drive
  hydrodynamic instabilities, as well as the steric interactions of
  rod-like particles that lead to nematic alignment. This active
  nematic system yields complex flows and disclination defect dynamics
  very similar to phenomenological Landau-deGennes $Q$-tensor theories
  for active nematic fluids, as well as by more complex Doi-Onsager
  theories for polar microtubule/motor-protein systems.  We apply the
  quasi-equilibrium Bingham closure, used to study suspensions of
  passive microscopic rods, to develop a non-standard $Q$-tensor
  theory. We demonstrate through simulation that this ``$BQ$-tensor''
  theory gives an excellent analytical and statistical accounting of
  the suspension's complex dynamics, at a far reduced computational
  cost. Finally, we apply the $BQ$-tensor model to study the dynamics
  of Extensor suspensions in circular and bi-concave domains. In
  circular domains, we reproduce previous results for systems with
  weak nematic alignment, but for strong alignment find novel dynamics
  with activity-controlled defect production and absorption at the
  boundaries of the domain. In bi-concave domains, a Fredericks-like
  transition occurs as the width of the neck connecting the two disks
  is varied.
\end{abstract}
\maketitle

\section{Introduction}

Active suspensions are non-equilibrium materials composed of suspended
particles whose activity, driven by consumption of a local fuel, lead
to particle motions or local induced flows \cite{ramaswamy10,SS2013}. Examples
include bacterial swarms \cite{dombrowski04,zhang10}, collections of
synthetic colloidal particles \cite{PalacciEtAl2013,bricard13}, and
mixtures of cytoskeletal filaments driven by molecular motors
\cite{SanchezEtAl2012,Shelley2016}. One origin
of the instabilities and complex dynamics in active suspension is the
stresses created in the surrounding solvent by the particles'
activity. The solvent is a coupling medium for multiscale dynamics:
particle interactions through the solvent can manifest at
the system scale as collective motion, which feeds back to alter their
interactions. Theoretical and numerical studies have investigated
various aspects of active suspensions at different scales, from single
particle dynamics to suspension rheology. While discrete-particle
simulations that incorporate long-ranged hydrodynamic effects can
capture observed large-scale features of these systems
\cite{hernandez-ortiz05,SS2007,SS2011,Yeo2015,Delmotte2015},
constructing accurate continuum models is essential to their
characterization and analysis, and for making new predictions of
macroscale behaviors \cite{SS2013}.

Continuum descriptions of active suspensions of self-propelled
particles, such as bacteria, have been based on a variety of
approaches
\cite{simha02,aranson07,underhill08,SS2008a,wolgemuth08,baskaran09,
  subramanian09}. In all, coarse-grained variables, such as particle
concentration, concentration-dependent steric interactions, order
parameters, and solvent stresses and velocity, are used to capture
salient features of the macroscopic dynamics. In our own work, we and
our collaborators have developed Doi-Onsager kinetic models to
describe the evolution of suspensions of active rod-like particles
\cite{SS2008a,SS2008b,HS2010,ESS2013,GBGBS2015a,GBGBS2015b}. In this
approach, a Smoluchowski equation describes the evolution of a
particle distribution function, coupled to a coarse-grained Stokes
equation driven by the extra stresses created by particle activity and
interactions. The fluxes and stresses of the model are derived from
modeling how such particles interact with mean-field flows and other
coarse-grained variables. For suspensions of self-propelled rods which
interact only hydrodynamically \cite{SS2008a,SS2008b} this approach
identifies characteristic aspects of the dynamics: either uniform
isotropic or globally aligned suspensions have hydrodynamic
instabilities that depend upon the self-propulsion mechanism, particle
concentration, and system size \cite{SS2008b,subramanian09,HS2010},
leading to the emergence of complex, turbulent-like dynamics. When
such particles also interact sterically through
concentration-dependent alignment torques, this generates a polar
``active-nematic'' model with an isotropic-to-nematic ordering
transition. The new nematically-aligned steady states can again be
unstable to hydrodynamic instabilities, again leading to complex
dynamics but modulated by an effective liquid crystal elasticity
\cite{ESS2013}.  Similar but more elaborate kinetic models have been
derived for ``bio-active'' suspensions \cite{SanchezEtAl2012} composed
of microtubules whose relative motions are driven by processive
cross-linking molecular motors such as kinesin-1
\cite{GBGBS2015a,GBGBS2015b}. There, microtubule motion and motor
activity create polarity-dependent fluxes and stresses that drive
hydrodynamic instability, and a complex dynamics characterized by
nematic director fields with motile disclination defects that are
continuously nucleated and annihilated.

While having the advantage of being founded upon microscopic modeling,
such kinetic theories can carry the price of complexity. For
Doi-Onsager theories as above, the distribution function depends upon
both particle position $\x$ and orientation vector $\p$ ($|\p|=1$) and
so, in three dimensions, has five independent variables plus time
$t$. This makes their simulation challenging. Hence, simplifications
have been sought, often following from moment-closure schemes that
eliminate the $\p$ variable, yielding more approximate models that
describe the evolution of lower-order $\p$-moments of the distribution
function, such as the concentration $\phi(\x,t)$ (zeroth-moment), the
polarity vector $\q(\x,t)$ (vector of first-moments), and the
unnormalized tensor order parameter $\D(\x,t)$ (tensor of
second-moments) \cite{woodhouse2012,SS2013,ezhilin2015,Theillard2017}.
Tensorial moment-closure models can resemble, with important
differences, tensor models based upon phenomenological Landau-deGennes
$Q$-tensor liquid crystal theories \cite{TGY2013,giomi13,Giomi2015}.

In this paper, we use a Doi-Onsager theory to study what is perhaps
the simplest active suspension: a collection of non-motile, yet
mobile, rod-like particles that exert active dipolar stresses on the
solvent and interact with each other through hydrodynamic coupling and
steric alignment torques. When the stresses are extensile, we call
these particles ``Extensors''. While this is a special case of the systems
above, it is well-worth studying in its own right. Firstly, it presents
all the basic transitions and instabilities associated with motile
suspensions but in a more revealing and simplified form. Secondly,
there are several types of active suspensions described at
least qualitatively by this theory, including particles that
elongate through stretching or growth, creating the extensile flows
associated with destabilizing dipolar stresses. Such particle dynamics
occurs in some instances of bacterial cell division \cite{Adams09}, as well as
in intriguing liquid-crystal phase transitions
\cite{BPR1987,SU2000}. Such stretching is also thought to occur for
bundles of microtubules (formed by a depletion interaction) as they
are ``polarity-sorted'' by kinesin-1 motor-proteins
\cite{SanchezEtAl2012,keber14}. A very different system is composed of
tripartite Au-Pt-Au nanomotors \cite{Wykes16,Jewell2016} for which
catalytic surface reactions with a surrounding aqueous hydrogen
peroxide solution \cite{paxton04} create extensile surface flows. The
self-assembly of motile aggregates by small numbers of such nonmotile
particles has been studied experimentally by Jewell {\it et al.}
\cite{Jewell2016} and Wykes {\it et al.}  \cite{Wykes16}, and
numerically by Pandey {\it et al.}  \cite{Pandey2016}.

Thirdly, we use this simple kinetic theory as a testing ground for constructing coarse-grained
macro models for active suspensions using the Bingham moment closure
\cite{Bingham74}, a closure scheme used successfully in the study of
passive rod suspensions \cite{Chaubal98,Feng98}. The Bingham closure
expresses fourth-moment tensors arising in the kinetic theory in terms
of $\D(\x,t)$ using the so-called Bingham distribution, which is a
quasi-equilibrium {\it Ansatz}. The Doi-Onsager system is then expressed
solely in terms of $\D$. We call this $BQ$-tensor theory as $\D$
is an unnormalized form of the tensor order parameter $\Q$, itself the
central evolved quantity in the Landau-deGennes theory of liquid
crystal dynamics, which has been applied extensively to so-called
active nematics \cite{TGY2013,giomi13,Giomi2015,GBGBS2015a,GBGBS2015b}. We show that the $BQ$-tensor
theory closely or exactly reproduces the instabilities of the kinetic
theory near homogeneous isotropic or nematic steady states. We then
show through simulations that the $BQ$-tensor theory gives an excellent
statistical accounting of the complex dynamics resolved by the kinetic
theory model, both with and without steric interactions, of an active
nematic suspension driven by extensile stresses.

Finally, having established the fidelity of the $BQ$-tensor theory, we
use it to examine the dynamics of active suspensions under
confinement. This is inspired by several recent experiments showing
that confined collective suspensions of self-propelled particles can
organize into auto-circulating states \cite{Wioland2013,wioland2016},
and develop emergent ordering and density shocks \cite{bricard13}. The
dynamics of active suspensions has been studied previously using both
discrete particle simulations and continuum models in straight
channels \cite{ezhilin2015,tsang2016}, circular disks
\cite{woodhouse2012,Lushi2014}, annuli, and racetracks
\cite{Theillard2017}. For suspensions in a circular chamber we find
perfect agreement with the previous results for suspensions that
interact only hydrodynamically \cite{woodhouse2012}, and focus instead
on the active nematic case where steric interactions are strong. We
observe a plethora of dynamical states as the activity is
increased. In all, defects are produced in the bulk, propagate, and
annihilate, both with each other and at the boundaries. In biconcave
geometries, which are essentially two disks smoothly connected by a
bridge, we find evidence for a Friedricks transition in which the
elasticity of the ordered fluid prevents flows between the two
sides. When the neck is wide enough, activity overcomes the
elasticity, and material (and defects) move between the two sides.

The paper is organized as follows. In Sect.~\ref{sec:model} we
develop the kinetic theory for microscopic rods that produce extensile
surface flows. Except for the sign of the active dipolar
stress, this theory is nearly identical to the classical Doi-Onsager
models for passive liquid crystal polymers where the suspended
molecules have fore-aft symmetry \cite{DE1986,Feng98}. We discuss its
analytical and stability properties, as well as introduce simplified
models, particularly coarse graining through the Bingham closure.  In
Sect.~\ref{sec:numerical}, we study numerically the dynamics of
Extensor suspensions. In periodic open domains, we implement a
pseudo-spectral method to solve both the kinetic theory and the
$BQ$-tensor theory, and compare the two descriptions in terms of
emergent coherent structures, flow patterns, and statistical
structure. We then simulate the dynamics of the $BQ$-tensor model in
both circular and biconcave domains using a finite element method.
Following a discussion in Sect.~\ref{sec:conclusion}, we present
some additional results in appendices.

\section{Doi-Onsager theory for a simple active suspension}
\label{sec:model}
\subsection{A micro-mechanical model}
\begin{figure}
\begin{center}
  \includegraphics[width=0.7 \textwidth]{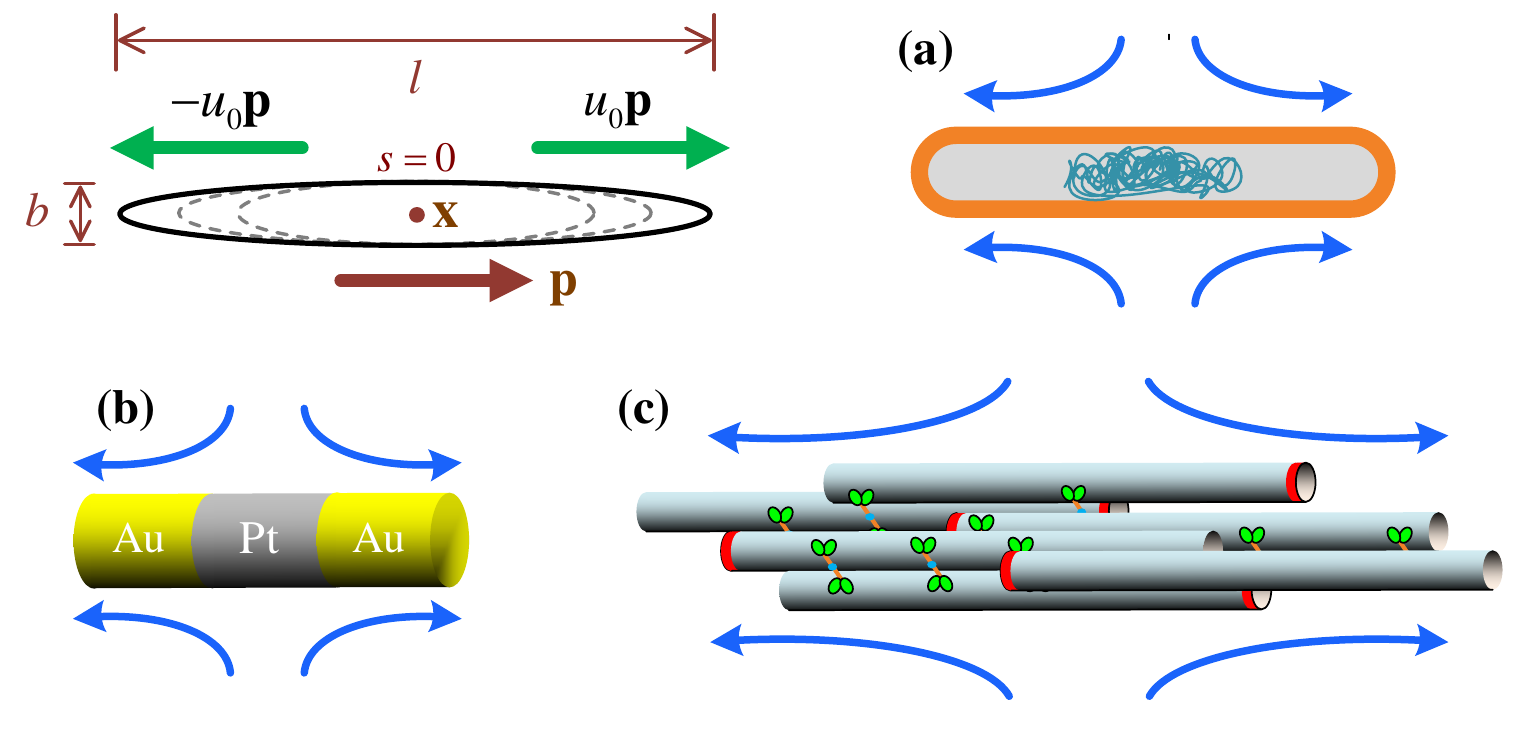}
\end{center}
\caption{(a) A schematic of an Extensor particle that produces extensile
  surface flows which yield extensile dipolar stresses. Related examples:
  (b) bacterial filamentation; (c) Au-Pt-Au nanomotor;
  (c) bundle of MTs mixed with motile crosslinking motors.}
  \label{fig:schematic}
\end{figure}
Consider a suspension of $N$ rigid rod-like active particles in a
volume $V$, each of length $l$ and diameter $b$. The rods are
considered to be slender so that the aspect ratio $r=b/l<<1$. For
definiteness we assume a simple form of activity: each rod
produces a symmetric surface flow $\v(s)=sgn(s)u_0\p$ with $-l/2\leq s
\leq l/2$ the signed arclength along the rod center-line, $\p$ its
unit orientation vector, and $u_0$ the signed surface flow speed; see
the schematic in Fig.~\ref{fig:schematic}. Because of surface flow
symmetry such active rods are not motile. If $u_0>0$ there is an
extensional straining flow along the rod, and a compressive one if
$u_0<0$. These flows are associated with dipolar extra
stresses. Directly following the derivation for Pusher suspensions
\cite{HS2011}, assuming that active particles are small relative to
the large-scale suspension flows they create {\it in ensemble}, and
using slender body theory for the Stokes equations, one can calculate
the rod center-of-mass velocity, its rate of rotation, and the
force/length exerted by the rod upon the fluid which, consequently,
gives the rod's contributions to the added stress. The single particle
contribution from the active surface flow is $\sigma_a \p\p$ where
$\sigma_a=-\eta l^2 u_0/8$ has units of force times length (i.e., a
``stresslet''), and $\eta=8\pi\mu/|\ln(er^2)|$ with $\mu$ the fluid
viscosity. Note that $\sigma_a<0$ for extensional surface flows, and
$\sigma_a>0$ for compressive ones. Overall, when $\sigma_a<0$ the
system can exhibit activity-driven hydrodynamic instabilities
\cite{simha02,SS2008a}.

\subsection{Doi-Onsager kinetic theory}
As an expression of the conservation of particle number, one can
derive a Smoluchowski equation for the (dimensional) distribution function
$\psi(\x,\p,t)$ for number density of particles at position $\x$ with
orientation $\p$:
\begin{equation}
  \label{eq:smol}
  \frac{\partial \psi}{\partial t} + \nabla \cdot (\dot{\x} \psi) +
  \nabla_{p} \cdot (\dot{\p}  \psi) =0.
\end{equation}
The distribution function is chosen so that $\int_{V} dV
\int_S dS_p \psi = N$.  Here and for the remainder of this
paper, derivative operators without subscripts (e.g., $\nabla$) denote
spatial derivatives. The orientational gradient operator on the unit
sphere is $\nabla_{p} = (\I - \p \p)\cdot\partial/\partial \p$.
Moments of $\psi$ with respect to $\p$ arise naturally in the theory:
the particle concentration $\phi$, (unnormalized) polarity field $\q$,
second-moment tensor $\D$ (unnormalized tensor-order parameter),
and fourth-moment tensor $\S$ are given by
\begin{eqnarray}
\label{Moments}
\phi(\x,t)&=&\int_S dS_{p} \ \psi(\x,\p,t),\\
\q(\x,t)&=&\int_S dS_{p} \ \p\ \psi(\x,\p,t) \\
\D(\x,t)&=&\int_S dS_{p}\ \p \p\ \psi(\x,\p,t),\\
\S(\x,t)&=&\int_S dS_{p} \ \p \p \p \p\ \psi(\x,\p,t).
\end{eqnarray}
The trace-free (normalized) tensor order parameter, the so-called
$Q$-tensor, is defined by $\Q(\x,t)=\phi(\x,t)^{-1}\D-\I/d$, with
$d=2,3$ the spatial dimension. The tensor $\Q$ has a
maximal nonnegative eigenvalue $\lambda_{max}$ satisfying
$0\leq\lambda_{max}\leq(d-1)/d$.  Assuming that $\lambda_{max}$ is
isolated, then we call its associated unit eigenvector the director
$\m$, and $0\leq s=\lambda_{max}d/(d-1)\leq 1$ the scalar order
parameter.

As described in previous work \cite{SS2008a,SS2008b,ESS2013,GBGBS2015a,GBGBS2015b}, the conformational fluxes $\dot\x$ and $\dotp$ arise from the dynamics of
the particle's center-of-mass position $\x$ and orientation $\p$,
calculated using slender-body theory for the Stokes equations
\cite{Keller1976}, and Maier-Saupe theory for steric alignment
\cite{maier58} of rod-like particles. This yields
\begin{eqnarray}
  \label{eq:force-balance}
  \dot{\x} &=& \u - D_T \nabla \ln \psi,\\
  \label{eq:jeffrey}
  \dotp &=& \left( \I - \p \p \right)\cdot (\gradu +2 \zeta_0
  \D)\cdot \p- D_R \nabla_{p} \ln \psi,
\end{eqnarray}
where $\u$ is the background fluid velocity field.
Equation~\eqref{eq:force-balance} states that the rod's positions move
with the background fluid flow and undergo translational diffusion
(with coefficient $D_T$), while Eq.~\eqref{eq:jeffrey} extends
Jeffrey's equation for the rotation of rods by the fluid velocity
gradient to include a local alignment torque arising from a rod-rod
interaction potential with strength $\zeta_0$. The last term of
Eq.~\eqref{eq:jeffrey} models rotational diffusion with coefficient
$D_R$.

The equations are closed by relating the incompressible background
velocity $\u$ and pressure $\Pi$ to the extra-stress tensor
$\bsig$ produced by the particles' presence \cite{batchelor70},
through a forced Stokes equation
\begin{eqnarray}
\label{eq:stokes}
\nabla \Pi - \mu \Delta \u &=& \nabla\cdot \bsig,\\
\nabla\cdot \u &=&0.
\end{eqnarray}
The stress tensor $\bsig$ has three contributions:
$\bsig=\bsig_a+\bsig_c+\bsig_s$. The dipolar active stress tensor
$\bsig_a=\sigma_a\D$ arises from the active surface flows generated by
particles. For Extensor particles, $\sigma_a$ is negative, and
positive for ``Contractors'' (see Fig.~\ref{fig:schematic}). The
so-called constraint stress tensor $\bsig_c=\sigma_c\S:\E$, where $\E
= 1/2(\nabla \u +\nabla \u^T)$ is the symmetric rate-of-strain tensor
and $\sigma_c = \eta l^3/24$ arises from particle rigidity
\cite{DE1986,ESS2013}. The stress tensor generated by steric
interactions is $\bsig_s = - \sigma_s (\D\cdot \D - \S : \D)$ with
$\sigma_s = \pi \eta l^3\zeta_0/3$ \cite{DE1986,ESS2013}.

Following \cite{ESS2013}, we introduce an effective volume fraction
$\nu = nbl^2$ where $n = N/V$ is the mean number density
\cite{DE1986,ESS2013}. To nondimensionalize, we choose a
characteristic length scale $l_c = b/\nu$, velocity scale $|u_0|$, and
stress scale $\mu |u_0|/l_c$. The distribution function is normalized
such that $(1/V)\int_\Omega dV\int_SdS_{p}~\Psi(\x,\p,t)=1$. Equation~
\eqref{eq:smol} retains its form, where the translational and rotational
fluxes are now
\begin{eqnarray}
\label{xflux}
\dot{\x} &=& \u - d_T \nabla \ln \Psi,\\
\label{pflux}
\dotp &=& \left( \I - \p \p \right) \cdot \left( \gradu
   + 2 \zeta \D\right)\cdot \p - d_R \nabla_{p} \ln \Psi.
\end{eqnarray}
The Stokes equation becomes
\begin{eqnarray}
\label{eq:stokes2}
\nabla \Pi - \Delta \u &=&  \nabla\cdot \bsig,\\
\label{eq:stokes2_2}
\nabla\cdot \u &=&0,
\end{eqnarray}
where the dimensionless stress tensor is given by
\begin{equation}
  \label{eq:stress2}
  \bsig = \alpha \D + \beta\S : \E - 2 \zeta \beta \left(\D\cdot \D - \S : \D  \right).
\end{equation}
The dimensionless control parameters are
\begin{equation}
  \label{eq:coeff}
  \alpha = \frac{\sigma_a}{\mu |u_0| l^2},\
  \zeta = \frac{\zeta_0}{|u_0| l^2} ,\
  \beta = \frac{\pi r\nu}{6 \ln(2r)},\
  d_T = \frac{\nu}{b |u_0|} D_T,\  d_R = \frac{b}{\nu |u_0|}D_R.
\end{equation}
Given the normalization of $\Psi$, the average density is
$\bar{\phi}=1$, and the state of uniform isotropy is given by
$\Psi\equiv\Psi_0 = 1/4\pi$ in 3D, and $1/2\pi$ in 2D. In the case of
spatially uniform density, $\D = \Q + \I/d$. Note that $\alpha$
inherits the sign of $\sigma_a$, and that the nondimensional
translational and rotational diffusion coefficients are proportional
and inversely proportional, respectively, to the effective volume
fraction $\nu$. We will sometimes refer to $\alpha$ as the activity
and treat it as a free parameter though, in fact, it is a geometric
parameter associated with particle shape and the relative placement of
active and passive stresses upon it \cite{SS2008a,HS2010,GBGBS2015a}.
For the specific case of the tripartite rods considered by Wykes {\it
  et al.}  \cite{Wykes16}, it is estimated that $\alpha\approx -1$.
Alternatively, if $u_0$ is interpreted as the velocity scale for a
base level of activity, then increases in $\alpha$ (either positively
or negatively) can be interpreted directly as multiplicative increases
in surface velocity over this base level.

Rotational thermal fluctuations of rod-like particles also give rise
to an extra stress of form $K\D$ with $K>0$ \cite{DE1986}. Our
assumption and expectation is that the deterministic active stress,
with $\alpha$ of either sign, is a much larger contribution.

A very similar model describes suspensions of stretching filaments
where the distribution function now depends upon particle length
$l(t)$. Assuming that particle length growth, $\dot{l}=f(l)$, is
independent of particle position and orientation, one can derive a
closed evolution equation for the marginal distribution by integrating out
$l$. The main differences are that $u_0$ is replaced by $f(l)$,
and $l$ dependencies are replaced by distributional averages. As
expected, $\dot{l}>0$ yields extensile stresses.

\medskip
Here we discuss some analytical properties of the kinetic model.

\smallskip
\noindent{\bf 1. Concentration Fluctuations.} Unlike suspensions of
Pusher swimmers, which also produce extensile flows and where
fluctuations can grow through nonlinear effects \cite{SS2008b,HS2011},
concentration fluctuations in the nonmotile case decay to zero. Denote
the fluid domain as $\Omega$, and assume either {\bf (i)} $\Omega$ is
a cubic domain $[-L/2,L/2]^3$ over which $\Psi$ is periodic, or {\bf
  (ii)} there is neither mass nor particle flux across the boundary
$\partial\Omega$. Integrating Eq.~(\ref{eq:smol}) over the unit
$\p$-sphere, and using Eq.~(\ref{eq:force-balance}) and velocity
incompressibility, then gives
\begin{equation}
\frac{\partial \phi}{\partial t}+\u\cdot\nabla \phi = d_T \Delta \phi.
\label{eq:concent}
\end{equation}
Integrating Eq.~\eqref{eq:concent} over $\Omega$, integrating by parts
and applying boundary conditions (periodic or zero flux) yields
\begin{equation}
\frac{d}{dt}\int_\Omega dV \frac{1}{2}\phi^2
= -d_T\int_\Omega dV |\nabla \phi|^2.
\end{equation}
Hence, concentration fluctuations from equilibrium will decay to zero,
regardless of particle type.

\smallskip
\noindent{\bf 2. An Energy Identity.} Like the motile case
\cite{SS2008b}, the kinetic theory has an ``energy'' identity
governing the evolution of a total energy composed of the
configurational entropy and the steric-interaction energy
\cite{Han15}. Let
\begin{equation}
{\cal E}={\cal S} + \kappa {\cal P} = \int_\Omega dV \int_S dS_{p}
\frac{\Psi}{\Psi_0}\ln \frac{\Psi}{\Psi_0}
- \kappa \int_\Omega dV ~\zeta (\D : \D) ,
\label{Energy}
\end{equation}
where $\kappa=d\beta/\alpha$. The conformational entropy ${\cal S}$
is non-negative, and zero if and only if $\Psi\equiv\Psi_0$ (the state
of uniform isotropy). The contribution ${\cal P}$ arises from the
Maier-Saupe approximation to the classical Onsager potential
\cite{Onsager1949}. We can show that ${\cal E}$ satisfies the identity
\begin{align} {\dot{\cal E}} = &- \frac{d}{\alpha} \left(
    2\int_\Omega dV ~\E:\E + \beta\int_\Omega dV \E:\S:\E
  \right) - \frac{4d\zeta^2 \beta}{\alpha} \int_\Omega
  dV~\D:\left(\D\cdot\D-\S:\D \right)
  \nonumber \\
  &+ 2d\zeta\left(1+\frac{2\beta d_R}{\alpha} \right) \int_\Omega
  dV ~\D:\D +\frac{d\zeta\beta d_T}{\alpha } \int_\Omega
  dV ~{\left| {\nabla {\bf{D}}} \right|^2}
  \nonumber \\
  &- \int_\Omega dV \int_S dS_p \left( d_T\left| \nabla\ln \Psi
    \right|^2 + d_R\left| \nabla_p\ln \Psi \right|^2 \right),
\label{EnergyDot}
\end{align}
where every integral density is non-negative (in particular,
$\D:(\D\cdot\D-\S:\D)\geq 0$). The first term in braces is a generalized
rate of viscous dissipation, the next three terms are generated by
steric interactions (and hence are multipled by $\zeta$), and
the last term is a negative-definite dissipation term arising from
thermal fluctuations.

Apparently, the above identity reveals that the nature of the
active stress, i.e., contractile or extensile, is a central
determinant of system behavior. For a contractile active stress
($\alpha>0$) in the absence of steric interactions ($\zeta=0$), the
energy $\mathcal{E}$ is driven to zero, corresponding to a uniform
isotropic state. The same result holds for the motile case
\cite{SS2008b} and is consistent with the simulations of semi-dilute active
particle suspensions \cite{SS2011}. When steric interactions are more
dominant, ordering effects can drive active contractile systems into
complex dynamics \cite{ESS2013}. When $\alpha<0$ (i.e., Extensor particles),
the particle activity increases the
energy in the system through the first term, while its effect upon the
steric terms is somewhat complex.

\smallskip
\noindent{\bf 3. The stability of uniform isotropic suspensions.}
The linear theory of Extensor suspensions has a simple and
evocative structure that is obscured in the motile case. We give below the
results of a stability calculation using the stream
function $\bPhi$ (i.e., $\u = \nabla \times \bPhi$ ), and the details are given in Appendix~\ref{AppendixB}.

For an isotropic steady state we have $\Psi^0=1/4\pi$, $\D^0=\I/3$,
$\u^0=\0$, and
$S^0_{ijkl}=\frac{1}{15}\left(\delta_{ik}\delta_{jl}+\delta_{il}\delta_{jk}
  +\delta_{ij}\delta_{kl}\right)$. Perturbing this steady state as
$\D=\I/3+\varepsilon {\bf{D'}},\;{\bf{S}} = {{\bf{S}}^0} + \varepsilon
{\bf{S'}},\;{\bf{u}} = \varepsilon{\bf{u'}}$, where $\varepsilon<<1$,
we find the 3D Smoluchowski equation
(\ref{eq:smol}) can be linearized at $O\left(\varepsilon\right)$ as \cite{SS2008a,HS2010,ESS2013}:
\begin{equation}
\label{eq:linSmol}
\frac{\partial \Psi'}{\partial  t}-\frac{3}{4 \pi} \p\p:\E'
=\frac{3\zeta}{2 \pi} \p\p:\D' + d_T \Delta \Psi'
+ d_R\nabla^2_{p}  \Psi'.
\end{equation}
The momentum-balance equation in (\ref{eq:stokes2},\ref{eq:stokes2_2}), together with Eq. (\ref{eq:stress2}), can be linearized as
\begin{eqnarray}
\label{eq:linconcstokes}
\nabla \Pi-\left(  1+\frac{\beta}{15}\right)  \Delta\u'
&=&\left( \alpha-\frac{2}{5}\beta\zeta\right) \nabla \cdot \D', \\
\nabla\cdot\u'&=& 0.
\end{eqnarray}
Rewriting Eq.~(\ref{eq:linconcstokes}) in terms of $\bPhi$ allows significant simplification. After some
calculation (see derivations in Appendix~\ref{AppendixB}), and defining
$\g = \nabla^4 \bPhi$, the linearized equations reduce to
\begin{equation}
  \label{eq:b}
  \frac{\partial \g}{\partial  t} =  \left(-C_1\left(\beta\right)\alpha
  +C_2\left(\beta\right) \zeta -6d_R\right) \g + d_T  \Delta \g,
\end{equation}
where $C_1(z)=(3/5)/(1+z/15))$ and $C_2(z)=(4/5)+(6z/25)/(1+z/15)$.
Thus, growth or decay rates of the plane-wave solutions are given by
\begin{equation}
\sigma=-C_1(\beta)\alpha+C_2(\beta) \zeta -6d_R -d_T k^2~.
\label{growth rates}
\end{equation}
The revealing aspect of this linear analysis is that the effect of
activity, steric interactions, and rotational diffusion all compete at
the level of scale-independent dissipation and growth. Steric
interactions associated with ordering drive the system away from
isotropy, which is abetted by activity if $\alpha<0$ (i.e. for
Extensor particles). Further, as is generic for active suspension
theories in the absence of an external scale
\cite{SS2008a,SS2013,GBGBS2015a,GBGBS2015b}, we find the fastest
growing mode occurs as $k\rightarrow 0^{+}$
\begin{equation}
  \sigma(k=0^+)=-\left( \frac{3}{\beta + 15} \right)\alpha
  +2\left( \frac{\beta + 6 }{\beta + 15} \right)\zeta-6d_R.
\end{equation}
Hence, for periodic boundary conditions imposed on sufficiently large
domains, the fastest growing scale is directly associated with the
domain's first periodic mode (which agrees well with our simulation
results in Sect.~\ref{sec:numerical}.)

As discussed below, another accessible steady state when steric
interactions are strong is the nematic state. We will turn to a
numerical treatment to study its stability except in the case of
sharply aligned suspensions.

\subsection{Reduced Models}

In this section we will discuss two useful reductions of the kinetic
theory. In the first, we investigate a special case of suspensions
where the active particles are strictly aligned at each point in
space. This is an exact reduction of the kinetic theory, and is useful
for constructing phenomenological models illustrating the
instabilities of aligned suspensions. Since dynamical simulation of the
kinetic theory is demanding given the number of independent
variable (e.g., three in space, two on the unit sphere, plus
time in 3D), we investigate the Bingham closure which has been successfully applied to passive rod suspensions. This is a coarse graining
that expresses the fourth-order $\S$ tensor in terms of the
second-order $\D$ tensor, leading to a so-called $Q$-tensor dynamics
which we term $BQ$-tensor theory. We find that it gives an excellent
accounting of the complex dynamics we observe in the kinetic theory,
and at a far lower computational cost by removing the orientational
degrees of freedom.

\smallskip
\subsubsection{Sharply aligned suspensions.}
\label{Sharp}
One analytically useful approximation is to assume negligible
translational and rotational diffusion, and that the suspension is
sharply aligned \cite{SS2008b}. That is, we write $\Psi(\x,\p,t) =
\barphi\delta[\p - \q(\x,t)]$ where $\delta$ is the Dirac delta
function on the unit sphere, and further assume that the concentration
is uniform, i.e., $\phi\equiv\barphi=1$. Integrating the Smoluchowski
equation~(\ref{eq:smol}) against $\p$, and making use of $\nabla \cdot
\u = 0$ and the identity $\int_S dS_p\ \p \nabla_p\cdot (\dotp \Psi) =
- \int_S dS_{p}\ \dotp \Psi$ \cite{SS2008a} yields
\begin{equation}
\frac{\partial \q}{\partial t} = -\u\cdot\nabla \q + (\I-\q\q) \cdot
\nabla \u \cdot \q.
\label{eq:aligngovern1}
\end{equation}
The extra stress in Eq.~(\ref{eq:stress2}) simplifies to
$\bsig=\alpha\q\q+\beta\q\q\q\q:\E$, and $\u$ is solved by
\begin{eqnarray}
\label{eq:aligngovern2}
\nabla \Pi - \Delta \u &=& \nabla\cdot\left(\alpha\q\q
 + \beta\q\q\q\q:\E\right), \\
 \label{eq:aligngovern2_2}
\nabla\cdot\u &=& 0.
\end{eqnarray}
Equations~(\ref{eq:aligngovern1})-(\ref{eq:aligngovern2_2}) are apolar, and
also invariant under the transformation $\q \leftrightarrow -\q$.

\medskip
Here we have a few comments:

\smallskip
\noindent{\bf 1.} Equations
(\ref{eq:aligngovern1})-(\ref{eq:aligngovern2_2}) provide a useful
comparison with the Bingham closure discussed below. On this point, we
can derive an equation for the evolution of the dipolar tensor
$\D=\q\q$, using that
\begin{equation}
  \frac{\partial}{\partial t} \left(\q\q\right)
  = - \left(  \u\cdot\nabla\right)  \left(  \q\q\right)
+\nabla\u \cdot \left(  \q\q\right)  +\left(  \q\q \right)\cdot
\nabla \u^{T} -2\left(  \q\q:\E \right)  \q\q.
\end{equation}
This evolution equation is
\begin{equation}
\D^{\nabla}+2\left(\D:\E \right)\D=\0,
\label{qqEvol1}
\end{equation}
where $\D^{\nabla}=\frac{\partial \D}{\partial t}
   +\u \cdot\nabla \D - \left(\nabla \u \cdot \D
   +\D \cdot \nabla \u ^{T} \right)$ is the upper-convected time derivative.
Therefore, it becomes a closed
evolution equation when taken together with the momentum balance
equation and incompressible constraint:
\begin{eqnarray}
\label{eq:aligngovern3}
\nabla \Pi - \Delta \u &=& \nabla\cdot\left(\alpha\D
+\beta\D(\D:\E) \right), \\
\label{eq:aligngovern3_2}
\nabla\cdot\u&=&0.
\end{eqnarray}

\smallskip
\noindent{\bf 2.} The sharply aligned dynamics are interesting to
consider in the Lagrangian frame of the background flow. As a
reminder, this is defined by the flow map $\X(\balp,t)$ from the initial
position $\balp$ (i.e., the Lagrangian variable) to the current position
$\X$ through
\begin{equation}
  \frac{\partial \X}{\partial t}(\balp,t)=\u(\X(\balp,t),t)
\end{equation}
where $\X(\balp,0)=\balp$. The associated deformation tensor is then defined as
$\F=\partial\X/\partial\balp$ ($F_{ij}=\partial X_i / \partial
\alpha_j$). When defining the field of unit vectors
$\r(\balp,t)=\frac{\F\cdot\q_{0}}{\left\vert \F\cdot\q_{0}\right\vert}
$, and using the identity $\frac{D \left\vert \F \cdot
  \q_{0}\right\vert}{D t} = \left\vert \F \cdot \q_{0}\right\vert
\left( \nabla\u : \r \r\right)$, it is then straightforward to show
that $\r$ satisfies Jeffery's equation:
\begin{equation}
\label{rEvolve}
\frac{\partial\r}{\partial t}  =
\left(  \I-\r \r \right) \cdot \nabla \u \cdot \r.
\end{equation}
Noting that $\F(\balp,0)\equiv\I$ and setting
$\q(\balp,0)=\q_0(\balp)$, then the uniqueness of solutions to initial
value problems gives immediately that, in the Lagrangian frame, the
solution to Eq.~(\ref{eq:aligngovern1}) is
\begin{equation}
\q(\balp,t)=\frac{\F \cdot \q_{0}}{\left\vert \F \cdot \q_{0}\right\vert }.
\label{qsolution}
\end{equation}
Equation~(\ref{qsolution}) is simply a normalized version of the {\it
  Result of Cauchy} that $\bomega=\F\cdot\bomega_0$ with $\bomega$ the
vorticity field evolved by the 3D incompressible Euler equations
\cite{Childress2009}.  Naturally, this also yields a Lagrangian
expression for the tensor $\D=\q\q$:
\begin{equation}
  \D =\frac{\left(\F \cdot \q_{0}\right)
    \left(\F \cdot \q_{0}\right)}{\left\vert \F \cdot \q_{0}\right\vert^2 }
  = \frac{\F \cdot {\D}_0 \cdot \F^{T}}
  {{\rm{tr}}\left(\F \cdot {\D}_0 \cdot \F^{T}\right) }.
\end{equation}
To be consistent with its origins, ${\D}_0$ must be initialized by the
dyadic product of a unit vector with itself.  These expressions can be
used to derive an integro-differential version in the Lagrangian frame
akin to the vorticity-based Lagrangian formulation of the
incompressible Euler equations \cite{Childress2009}.

\smallskip
\noindent{\bf 3.} The sharply aligned dynamics, described either by
Eqs.~(\ref{eq:aligngovern1})-(\ref{eq:aligngovern2_2}) or
Eqs.~(\ref{qqEvol1})-(\ref{eq:aligngovern3_2}), is particularly useful
for analyzing the stability of aligned suspensions. A globally-aligned
steady state is given by $\q\equiv\zhat$ (or $\D=\zhat\zhat$) and
$\u\equiv\0$. Following \cite{SS2008b} by linearizing about this
state, and seeking plane-wave solutions with wave-vector $\k$, yields
two growth rates:
\begin{equation}
  \sigma_{1,2}=-\alpha H_{1,2}(\Phi),
  \label{LinearGrowthAligned}
\end{equation}
where $H_1(\Phi)=\cos^2\Phi$ and
$H_2(\Phi)=\cos^2\Phi\left(2\cos^2\Phi-1\right)$, with
$\Phi\in[0,\pi)$ the angle between the normalized wave-vector $\khat$
  and $\zhat$ (i.e., $\cos\Phi=\khat\cdot\zhat$). For $\alpha<0$,
  there is exponential growth at all plane-wave angles, independently
  of $k$, with the exception of $\Phi=\pi/2$ which is neutrally
  stable. On both branches, the direction of maximal growth is
  $\Phi=0$ for which the wave-vector is aligned with the
  suspension. This is in agreement with the more complex expression
  for motile suspensions, and also seems to be a generic result for
  active matter theories \cite{SS2008b,giomi2011,SS2013,GBGBS2015a}.
Maximal growth rates for any plane-wave vector $\k$ are again,
like the isotropic case, obtained as $k$ tends to zero.

Spatial diffusion can be incorporated in the model phenomenologically
by including the term $d_T(\I-\q\q)\cdot \Delta\q$ on the
right-hand-side of Eq.~(\ref{eq:aligngovern1}) \cite{lin1995}. This simply yields the
modified growth rates
\begin{equation}
  \tilde{\sigma}_{1,2}=-\alpha H_{1,2}(\Phi)-d_Tk^2.
  \label{ModifiedLinearGrowthAligned}
\end{equation}

\smallskip

\subsubsection{Coarse-graining via the Bingham closure}
\label{sec:Bingham}
The kinetic theory is high dimensional, having three spatial
dimensions and two angles on the unit sphere (in 3D), and thus
expensive to simulate. Alternatively, we can consider
approximations based on moment-closure schemes to find approximate
evolution equations for low-order $\p$-averaged quantities. Many such
schemes have been proposed \cite{DE1986,Feng98} but we consider here
the Bingham closure which is based on a quasi-equilibrium {\it Ansatz}
and has been previously applied and compared to kinetic theories
of passive rod suspensions in simple rheological flows \cite{Feng98}
(essentially the same as ours but with $\alpha>0$ and externally
forced). The zeroth moment with respect to $\p$ of the Smoluchowski
equation yields the concentration equation (\ref{eq:concent}), while
integrating the tensor $\p\p$ against the Smoluchowski equation yields
a dynamical equation for $\D$:
\begin{equation}
  \D^{\nabla}+2\E:\S
  =4{\zeta}\left( \D\cdot \D - \S : \D  \right)
  + {d_T}{\Delta}\D
  - 2d d_R\left( {\D
  - \frac{\phi}{d}{\bf{I}}} \right).
\label{eq:qtensor}
\end{equation}
The left-hand-side of Eq.~(\ref{eq:qtensor}) is a tensor transport
operator with its last term being generated by Jeffery's
equation \cite{Jeffery22}. This evolution equation for $\D$ will be closed if the
fourth-moment tensor $\S$ is expressed in terms of lower order
moments, in particular if $\S=\S[\D]$. Here we use the
Bingham closure \cite{Bingham74,Chaubal98} which constructs
an approximate local orientation distribution function $\Psi_B$
given the second-moment tensor $\D$. The Bingham distribution takes
the axisymmetric form:
\begin{equation}
{\Psi_B}[\T] = \frac{\exp \left(\T:\p\p \right)}{Z[\T]},
\label{eq:bingham}
\end{equation}
where $\T$ is a traceless symmetric tensor and $Z$ is a normalization
constant. For given $\D$, the tensor $\T$ is determined by solving the relation
\begin{equation}
\D =\int_S dS_p \p\p \Psi_B[\T]
\end{equation}
at each point in space. Computationally, one can take advantage of the
fact that the eigenvectors of $\T$ and $\D$ are co-aligned and so both
can be rotated into diagonal forms in a common frame. Consequently,
only the eigenvalues of $\T$ need to be determined
\cite{Chaubal98}. Given $\T$, and hence $\Psi_B$, the fourth-moment
tensor ${\bf S}$ is then approximated by $\S_B[\D]=\int_S dS_p
{\Psi_B}\p\p\p\p$. The final equations take the following forms:
\begin{equation}
\D^{\nabla}+2\E:\S_B[\D]
  =4{\zeta}\left( \D\cdot \D - \S_B[\D]: \D  \right)
  + {d_T}{\Delta}\D
  - 2d d_R\left( {\D
    - \frac{\phi}{d}{\bf{I}}} \right),
\label{Bingham1}
\end{equation}
together with
\begin{eqnarray}
\label{Bingham2}
 \nabla \Pi - \Delta \u &=& \nabla\cdot \bsig_B[\D], \\
\label{Bingham2_2}
 \nabla\cdot \u &=& 0.
\end{eqnarray}
And the extra stress is expressed as:
\begin{equation}
  \label{Bingham3}
 \bsig_B[\D]=\alpha\D+\beta\S_B[\D]:\E
  - 2\zeta \beta\left(\D\cdot\D -\S_B[\D]:\D \right).
\end{equation}
We refer to Eqs.~(\ref{Bingham1})-(\ref{Bingham3}) as the
$BQ$-tensor model. We can also show that both Eq.~(\ref{eq:qtensor}) and  Eq.~(\ref{Bingham1}) conserve $\rm{tr}(\D)$.

\medskip
For the above $BQ-$tensor model, we have several comments:

\smallskip
\noindent {\bf 1.} Previous work \cite{Chaubal98,Feng98} has shown
that in potential flows and in the weak flow limit (i.e., the
equilibrium state), the Bingham distribution function
yields the exact solution of the Smoluchowski equation (with the
rotational flux only). The Bingham distribution also arises naturally
as describing spatially uniform, nematically-ordered steady states.
Such steady states can be obtained by setting $\dotp=\0$ in
Eq.~(\ref{pflux}), leading to a balance between the rotational diffusion and
the Maier-Saupe alignment torque (see, for example, Ezhilan {\it et
  al.} \cite{ESS2013} and Gao {\it et al.} \cite{GBGBS2015b}):
\begin{equation}
  {\nabla_p}\ln {\Psi} = \xi \left( {{\bf{I}} - {\bf{pp}}}
  \right)\cdot {\D}\cdot {\bf{p}},
\label{eq:bingham3d_1}
\end{equation}
where $\xi = 2\zeta/d_R$. Consider the unit sphere with angle
coordinates $(\theta,\Phi)$ and polar axis along $\hat{\bf z}$, so
that $\p = \sin\Phi \cos\theta \hat{\x} +\sin\Phi \sin \theta
\hat{\bf{y}} +\cos\Phi \hat{\bf{z}}$ with $\theta\in[0,2\pi)$ and
$\Phi\in[0,\pi)$. In seeking a solution $\Psi(\Phi)$, we find $\D$ has the
form
\begin{equation}
\D[\Psi] = A[\Psi]{\bf{\hat z\hat z}}+B[\Psi]\I
\end{equation}
where $A[\Psi]=\pi\int_0^\pi d\Phi\Psi(3\cos^2\Phi-1)\sin\Phi$ (see
\cite{ESS2013}). Then Eq.~(\ref{eq:bingham3d_1}) can be integrated to
\begin{equation} \Psi(\Phi) = \frac{\exp\left(\delta(\xi)\cos 2\Phi\right)}
{2\pi\int_0^\pi\exp\left(\delta(\xi)\cos 2\Phi\right)\sin\Phi d\Phi}
= \frac{\exp(\T:\p\p^T)}{Z[\T]},
\label{eq:bingham3d_2}
\end{equation}
where ${\bf{T}} =\delta(\xi) {\rm{diag}}\left(-2/3,-2/3,4/3 \right)$.
In the above, $\delta(\xi)$ is the solution of the equation:
\begin{equation}
\delta = \frac{\xi}{8}
\frac{\int_0^\pi d\Phi\sin\Phi (3\cos^2\Phi - 1)\exp\left(\delta \cos 2\Phi\right)}
{\int_0^\pi d\Phi\sin\Phi\exp\left(\delta\cos 2\Phi\right) },
\label{eq:nematic}
\end{equation}
which captures a bifurcation, with increasing $\xi$, from an isotropic
($\delta(\xi)=0$, $\Psi\equiv 1/4\pi$, and $A[\Psi]=0$) to a
nematically aligned state ($\delta(\xi)\neq0$ and $A\neq 0$).

Thus the Bingham distribution has a microscopic origin with respect to
the kinetic theory, and captures the the isotropic-to-nematic
transition.  However, to our knowledge there is no asymptotic
analysis, rigorous or formal, that establishes error estimates between
the Bingham closure and the original kinetic theory. We suspect that
such a connection could be established in the particular limit of
strong alignment torques at large $\xi$.


\smallskip
\noindent{\bf 2.} The $BQ$-tensor system is a direct closure of the
kinetic theory for an active or passive liquid-crystal
polymer suspension, and each of its parameters has a clear origin (though admittedly, the kinetic theory itself is
derived under several modeling and separation-of-scale assumptions).
While structurally similar to the active and passive nematic $Q$-tensor
theories that follow the phenomenological Landau-deGennes (LdG)
approach for liquid-crystalline fluids (see, for example,
\cite{Giomi2015}), there are significant differences. In the LdG
approach, the ordering and relaxation dynamics, and elastic stresses,
follow from the LdG free-energy. The $BQ$-tensor theory has different
nonlinearities governing the steric ordering, in both the $\D$ dynamics
and the associated stress, especially due to the presence of
$\S_B$. While in the LdG theory the spatial diffusion coefficient of
$\Q$ is a Frank elastic coefficient, in the $BQ$-tensor theory for
$\D$ it is the spatial diffusion coefficient, $d_T$, of the
active particles. In the LdG theory, the transport operator for $\Q$ is
corotational, while in the $BQ$-tensor theory it is, in part, an upper-convected
derivative arising from the microscopic modeling.  Finally, in the $BQ$-tensor
theory there are additional terms, in both dynamics and stresses, that arise
from the constraint of particle rigidity.

\smallskip
\noindent{\bf 3.} Note that while the original kinetic theory has an
energy law (i.e., Eqs.~(\ref{Energy},\ref{EnergyDot})), the $BQ$-tensor theory apparently does not. However, Li {\it et al.} \cite{LWZ2015} have recently developed more complicated Bingham closure models
for passive rod suspensions that preserve the energetic structure. In
their theory, the configurational entropy is approximated using the
Bingham distribution, leading to a theory where the $\T$ tensor
appears directly in the dynamics, rather than only indirectly in the
calculation of $\S_B$. This approach leads to a closer concordance of
the $BQ$-tensor and LdG $Q$-tensor theories, though we have not
pursued that here in the context of active suspensions.

\smallskip
\noindent{\bf 4.} In the absence of spatial diffusion,
Eq.~(\ref{Bingham1}) evolves $\D$ using only local information. Hence,
like the sharply aligned case, this equation could be reformulated
naturally to describe the dynamics in the Lagrangian frame.

\smallskip
\noindent{\bf 5.} One simple and direct comparison of the kinetic and
$BQ$ theories is their linear behavior near steady state. We examine
this in order, first for the uniform isotropic steady state, and second
for a nematically aligned steady state.

\smallskip
{\bf i.}~{\it Stability near uniform isotropy.} We will
show that the kinetic and $BQ$ theories give identical results.
This is easily seen by considering
the evolution of the perturbation $\D'$. Multiplying the linearized
Smoluchowski equation (\ref{eq:linSmol}) by $\p\p$, and integrating
over the unit $\p$-sphere gives the evolution equation for $\D'$:
\begin{equation}
\frac{{\partial {\bf{D'}}}}{{\partial t}} -\frac{2}{5} \E'
    = \frac{4\zeta}{5}{\bf{D'}}+ {d_T}{\Delta}{\bf{D'}}
    - 6d_R{\bf{D'}}.
\label{eq:Dprime}
\end{equation}
The above equation is closed since $\E'$ is a linear functional of
$\D'$ through the linearized Stokes equations
(\ref{eq:linconcstokes}). Inverting Eq.~(\ref{eq:linconcstokes}) in
Fourier space for $\u'$ (with $\u'(\x)=\tilde\u(\k)\exp(i\k\cdot\x)$),
forming $\tilde\E'[\tilde\D']$, and substituting into
Eq. (\ref{eq:linSmol}) yields
\begin{equation}
\tilde\D_t'=-\gamma\left( ( \I - \hat\k \hat\k) \cdot
\tilde\D' \cdot \hat\k \hat\k + \hat\k \hat\k
\cdot \tilde\D'\cdot(\I-\hat\k \hat\k) \right)
-\omega\tilde\D'.
\label{eq:app}
\end{equation}
where \begin{equation}
\gamma=\frac{1}{5}\left( {\frac{\alpha - \frac{2}{5}\zeta\beta}
{1+\frac{\beta}{15}}}\right)
\;\;{\rm{and}}\;\;
\omega_k=-\left(k^2d_T - \frac{4\zeta}{5} + 6d_R \right),
\label{LinearD1}
\end{equation}
and $\hat\k=\k/k$.

Linearizing the $BQ$-theory Eqs.~(\ref{Bingham1})-(\ref{Bingham2_2})
gives an identical result, which is mainly due to the fact that the
term $\D\cdot\D-\S:\D$ appears commonly in both the $\D$ transport and
momentum-balance equations (here we suppress the $B$ subscript in the
Bingham case). Also linearization of $\S:\D$ produces
$\S^0:\D'+\S':\D^0$, and in both models we have $\S^0:\D'=2\D'/15$ and
$\S':\D^0=\D'/3$.

To finish the analysis, we employ $\R$ as a rotation matrix such that
$\hat\k=\R \cdot\hat\z$, and define $\hat\D'=\R^T\cdot\tilde\D'\cdot\R$. Then
Eq.~(\ref{eq:app}) can be rewritten as
\begin{equation}
\hat\D_t'=-\gamma\left( \left( \I - \hat\z \hat\z \right) \cdot
\hat\D' \cdot \hat\z \hat\z + \hat\z \hat\z
\cdot \hat\D'\cdot\left(\I-\hat\z \hat\z \right) \right)
-\omega_k\hat\D'.
\label{eq:app2}
\end{equation}
In the rotated frame, the off-diagonal shear
components $\hat D_{13}'$ and $\hat D_{23}'$ evolve as
\begin{equation}
\hat D_t'=-(\gamma+\omega_k) \hat D',
\label{LinearD2}
\end{equation}
which yields the growth rate, $\sigma=-(\gamma+\omega_k)$, identical to that in Eq.~(\ref{growth rates}). The remaining components of
$\hat\D'$ instead satisfy
\begin{equation}
\hat D_t'=-\omega_k\hat D'.
\end{equation}
In short, only $\hat D_{13}'$ and $\hat D_{23}'$ reflect the
contributions of fluid flows, which can be stabilizing or
destabilizing. The remainder are decoupled from flow and show
decay or growth only due to thermal fluctuations and steric
interactions, respectively.

\smallskip
{\bf ii.}~{\it Stability near a nematically aligned homogeneous
  state.}  As discussed above, with the inclusion of steric
interactions via the Maier-Saupe potential there are additional
homogeneous solutions that arise as a balance between the steric
alignment torque and the rotational diffusion. These solutions are
parametrized by $\xi=2\zeta/d_R$, and only the isotropic solutions
exist when $\xi$ is below a certain critical value $\xi_c$. In 3D
there is a transcritical bifurcation at $\xi=\xi_c\approx 13.46$
beyond which the isotropic state becomes energetically unstable (in
terms of the Maier-Saupe interaction energy). The system will be
driven towards a nematic state where particles become increasingly
aligned as $\xi\rightarrow\infty$ \cite{ESS2013}. In 2D the
bifurcation is instead a supercritical pitchfork bifurcation occurring
at $\xi_c=8$ \cite{GBGBS2015b}. As discussed in Comment ({\bf 1.}) the
form of distribution function for the nematically aligned states is of
Bingham type, and hence yields steady states for the $BQ$-tensor
theory.


The plane-wave stability of nematically aligned states has been
examined in kinetic theories for Pusher and Extensor suspensions
\cite{ESS2013}, and for microtubule/motor-protein suspensions
\cite{GBGBS2015a,GBGBS2015b}. These show that activity can drive a
bending instability where the wave-vector of maximal growth is aligned
with the base orientation of the nematic state, as already suggested
by the sharp-alignment analysis (see
Eq.~(\ref{ModifiedLinearGrowthAligned})).  Here we examine the bending
instability for Extensor suspensions in 2D for both the kinetic and
$BQ$-tensor theories. Solving Eq.~(\ref{eq:nematic}) for $\delta(\xi)$
gives the Bingham distribution $\Psi_0$, the base-state tensors $\T_0$,
$\D_0$ and $\S_0$, and the base-state normalization factor $Z_0$. Perturbations
of these base quantities are introduced as ($\varepsilon \ll 1$)
\beq
{\bf{T}} = {{\bf{T}}^0} + \varepsilon {\bf{T'}}, \;\; Z = {Z^0} +
\varepsilon Z', \;\; {\bf{D}} = {{\bf{D}}^0} + \varepsilon {\bf{D'}},
\;\; {\bf{S}} = {{\bf{S}}^0} + \varepsilon {\bf{S'}}.
\eeq
The perturbations can be further expressed in terms of $\T'$ as
\begin{equation}
Z'={Z_0}{{\bf{D}}_0}:{\bf{T'}},\;\;\;
\D'=\left({{\bf{S}}_0} - {{\bf{D}}_0}{{\bf{D}}_0}\right):{\bf{T'}},\;\;\;
\S'=\left({{\bf{R}}_0} - {{\bf{S}}_0}{{\bf{D}}_0}\right):{\bf{T'}},
\end{equation}
with ${{\bf{R}}^0} = \frac{1}{Z^0}\int {{\bf{pppppp}}\exp \left(
  {{{\bf{T}}^0}:{\bf{pp}}^T} \right)} d{\bf{p}}$ the sixth-order
$\p$-moment tensor. These relations allow us to express the high-order
moments in terms of $\D'$ through the intermediate variable $\T'$. The
linearized equations are then transformed to Fourier space where we solve
for the plane-wave growth rates $\sigma$ numerically.

\begin{figure}
\begin{center}
  \includegraphics[width=0.5 \textwidth]{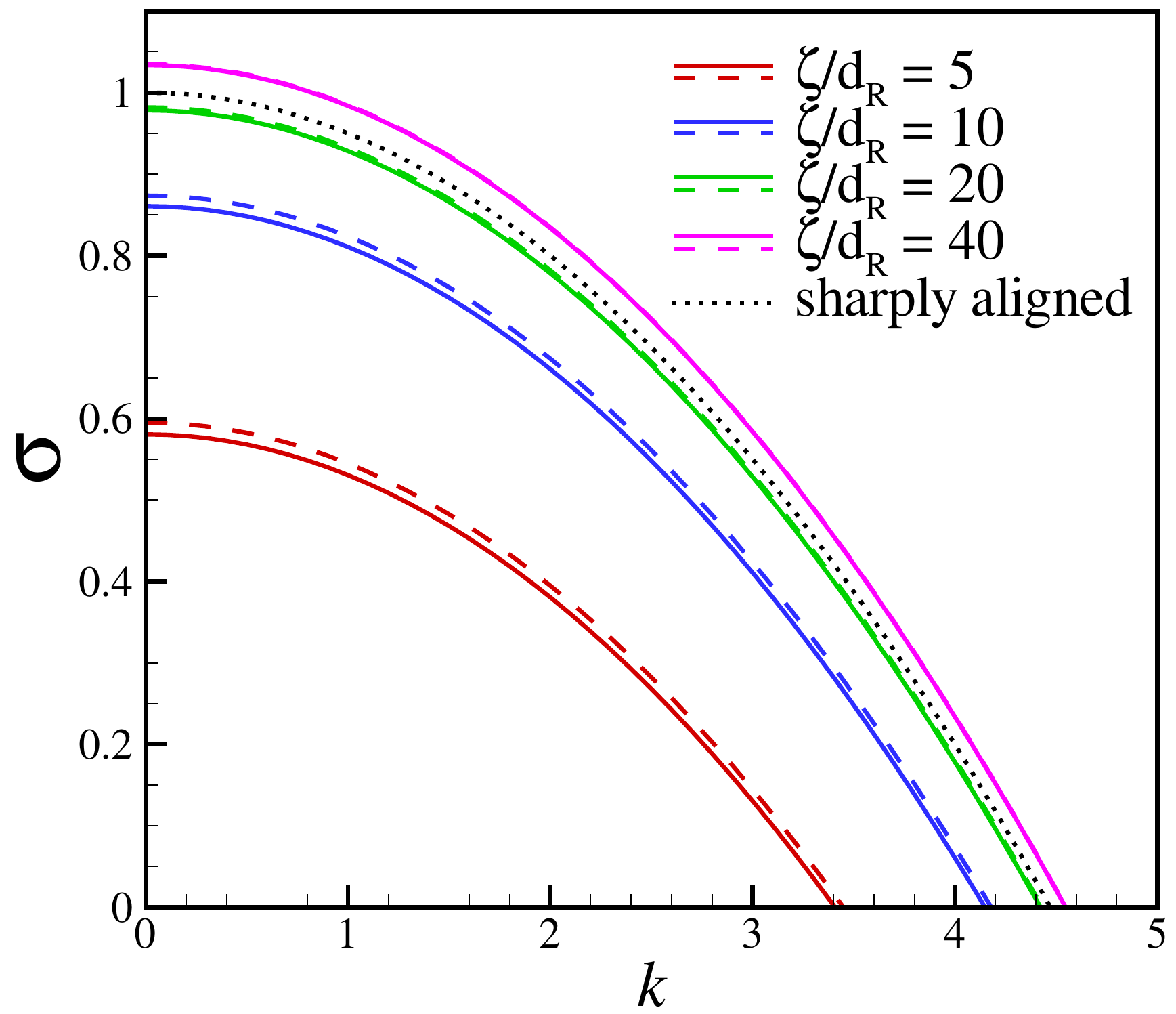}
\end{center}
\caption{growth rates as a function of wave-number $k$ as alignment
  torques are increased (i.e., increasing $\zeta/d_R$). The black
  dotted curve is the growth rate predicted from the sharply-aligned case. Here
  $\alpha=-1.0$, $\beta = 0.874$, and $d_T=d_R=0.05$.}
  \label{fig:regular}
\end{figure}

Figure \ref{fig:regular} shows the growth rates $\sigma$, for both the
kinetic and $BQ$-tensor theories, of a plane-wave perturbation to the
homogeneous nematic state. Here the plane-wave vector is in the
direction of nematic alignment, which has the greatest
growth rates. Similar to the isotropic case,  the fastest growth rate is
obtained as $k\rightarrow 0^+$.  These features are in agreement with
the sharply aligned analysis, and for motile and microtubule
suspensions.  We show the effect of increasing alignment torque
while also comparing the results of the kinetic theory with those of
the $BQ$-tensor theory.  As $\zeta$ is increased for fixed $\alpha$,
we see an increase in the maximal growth rate and the band of unstable
modes. We also see an increasingly good correspondence between the
full kinetic theory and the $BQ$-theory. In addition, when examining the prediction
from the sharply-aligned analysis (black-dotted curve), we find while it gives
an excellent accounting for the overall scale and features of the
growth rate curves, it is apparently not the asymptotic limit of the
$BQ$-tensor model as $\zeta$ tends to infinity.

\section{Numerical studies and comparisons}
\label{sec:numerical}

In this section we study numerically the dynamics of Extensor
suspensions in both periodic and closed geometries.  We use
simulations in 2D periodic geometries to compare the kinetic model
with its coarse-grained $BQ$-tensor model at different levels of activity (i.e., varying $\alpha$). Here we consider
two versions of the model (both kinetic and its $BQ$-tensor closure):
the full model, termed Model II, and Model I where we omit both steric
interactions and constraint stresses (i.e., set
$\zeta=\beta=0$). Model I makes a connection with our earlier studies
of active suspensions \cite{SS2008a,SS2008b}, and emphasizes the
fundamental differences arising from steric interactions and nematic
ordering. In either case we find that the $BQ$-tensor theory gives an excellent statistical accounting of the complex dynamics, particularly
for Model II. We then use the $BQ$-tensor theory to study the dynamics
of active suspensions under confinement. Flows in a disk have been
studied previously using other closures of Model I to show
bifurcations, with increasing activity, from isotropic to
auto-circulating flows, and thence to more complex dynamics.  We
recapitulate this using a $BQ$-tensor version of Model I, and find excellent agreement (see Appendix~\ref{AppendixC} for details). We then focus on the full model (i.e., Model II) to study confined suspensions when steric interactions are strong. These steric interactions make a
nematically ordered suspension the natural state against which
activity competes, and we find both new bifurcations and dynamical
phenomena.

\subsection{Numerical comparison of the kinetic and $BQ$-tensor theories}
In 2D, we evolve $\Psi(\x,\theta)$ where $\theta\in[0,2\pi)$ is the
  particle orientation in the $xy$-plane (i.e.,
  $\p=(\cos\theta,\sin\theta)$). For the kinetic theory we use
  Eqs.~(\ref{eq:smol},\ref{xflux},\ref{pflux}) to
  evolve the distribution function $\Psi$, with the requisite
  background velocity field obtained by solving Eq.~\eqref{eq:stokes}.
  The numerical method is pseudo-spectral with 256-512 Fourier modes
  in each spatial direction, and 64 modes in the orientation
  $\theta$. Equation \eqref{eq:stokes} is inverted via Fourier
  transform as
\begin{equation}
\tilde{\u}(\k) = \frac{i}{k^{2}}\left(\I-\hat{\k}\hat{\k}\right) \cdot
  \tilde{\bsig}  \cdot \k.
\label{eq:regular}
\end{equation}
We use pseudo-spectral collocation to evaluate the nonlinear advection
terms. For time-stepping, we use a second-order Adams-Bashforth scheme
of the Fourier-transformed distribution function, combined with an
integrating factor method for handling the diffusion terms accurately
and stably \cite{HLS1994,GBGBS2015a}.  The nondimensional diffusion
coefficients, $d_T$ and $d_R$, are chosen between 0.01-0.05.
We evolve the $BQ-$tensor system in a similar fashion, again using
256-512 Fourier modes in each spatial direction.

\begin{figure}
\begin{center}
  \includegraphics[width=0.8 \textwidth]{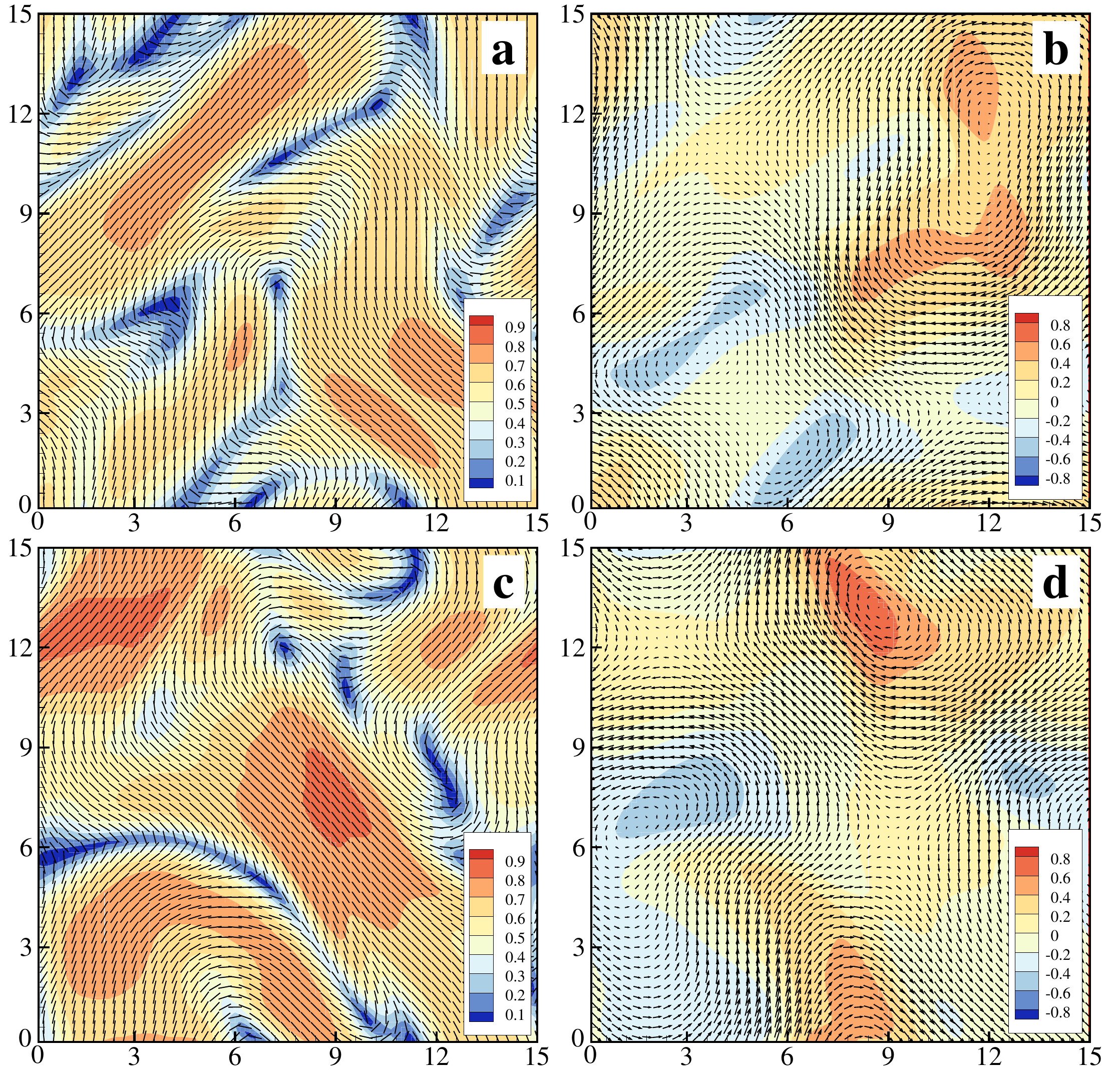}
\end{center}
\caption{Collective motion of Extensor suspensions at late
  times ($t=100$) when using Model I, starting from an initial condition of a small
  perturbation about the uniform state by kinetic theory (a,b) and
  $BQ-$tensor theory (c,d). Panels (a,c) show the nematic director
  field, superimposed on the colormap of the scalar order parameter
  $s$; right panels (b,d) show the background fluid velocity vector
  field superposed upon the colormap of the associated vorticity.  The
  simulation parameters are chosen as $L=15$, $\alpha=-1.0$,
  $d_T=d_R=0.02$. }
  \label{fig:evolution}
\end{figure}

\begin{figure}
\begin{center}
  \includegraphics[width=0.8 \textwidth]{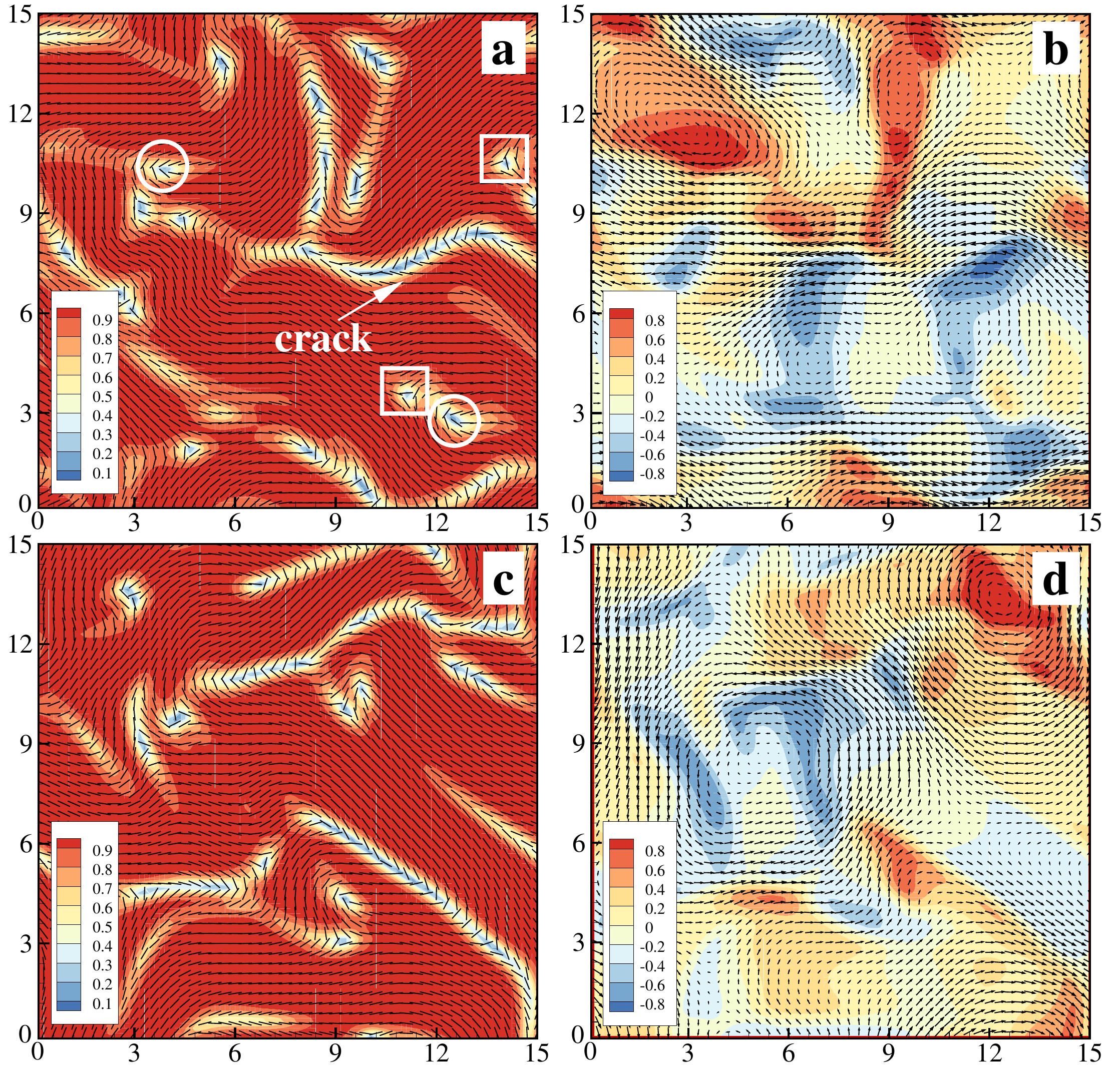}
\end{center}
\caption{Active nematic states of Extensor suspensions at
  late times ($t=100$) when using Model II, starting from near-uniform state by kinetic
  theory (a,b) and $BQ-$tensor theory (c,d). Panels (a,c) show the
  nematic director field, superimposed on the colormap of the scalar
  order parameter $s$; right
  panels (b,d) show the background fluid velocity vector field
  superposed upon the colormap of the associated vorticity. In (a),
  typical disclination defects are marked by open circles ($+1/2$) and
  squares ($-1/2$). The simulation parameters are chosen as $L=15$,
  $\alpha=-1.0$, $\zeta = 1.0$, $\beta = 0.874$, and $d_T=d_R=0.05$. }
  \label{fig:evolution_dense}
\end{figure}

\begin{figure}
\begin{center}
  \includegraphics[width=0.5 \textwidth]{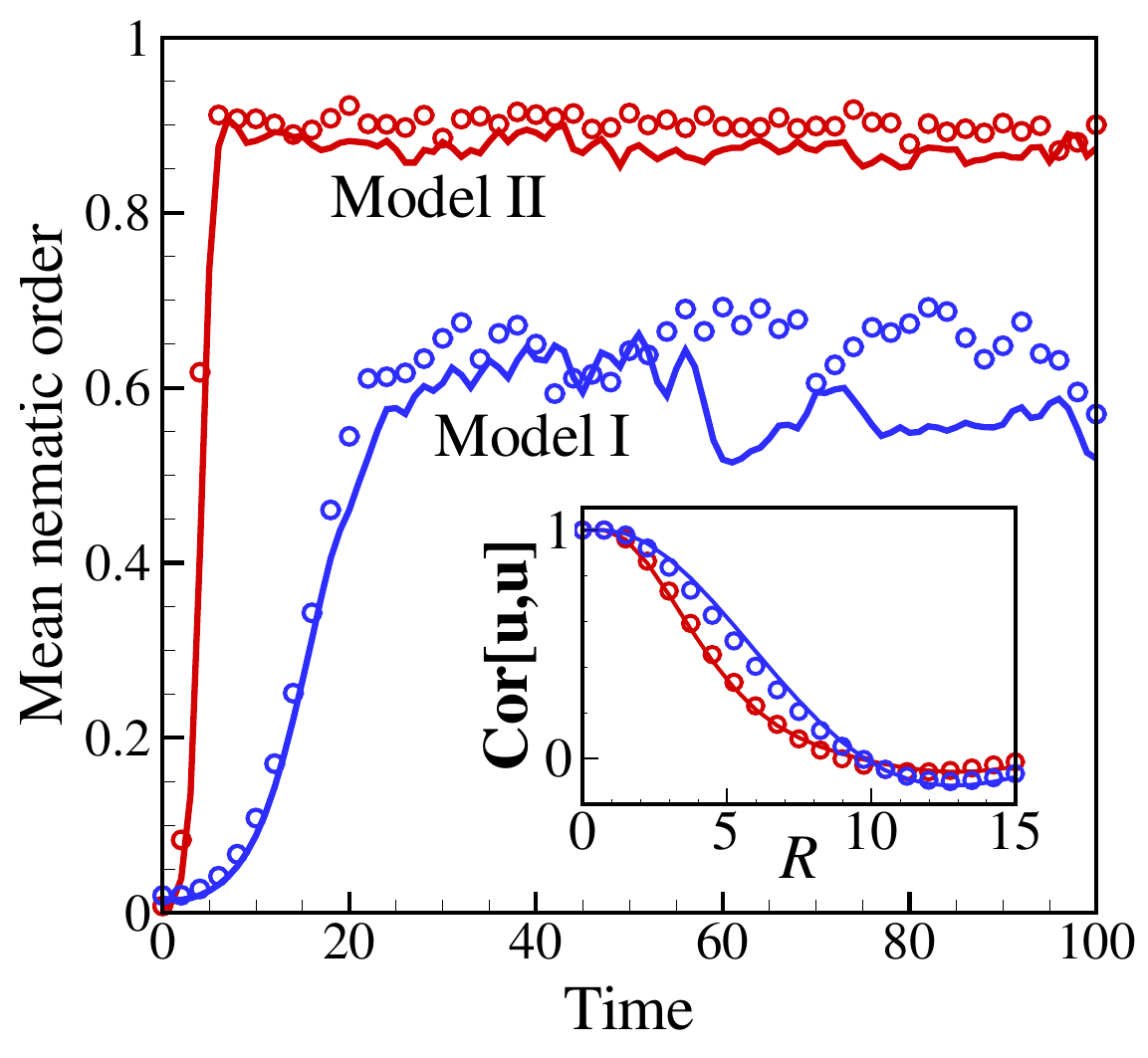}
\end{center}
\caption{The mean (i.e., spatially averaged) nematic order $S$ as a
  function of time. Inset: $\rm{Cor}\left[\u,\u\right]$. Solid lines and open circles represent the results of
  the kinetic theory and the $BQ-$tensor model, respectively.}
  \label{fig:lines}
\end{figure}



As a first comparison, Fig.~\ref{fig:evolution} shows the late-time
evolution of Model I for both the kinetic theory and its $BQ$-tensor
closure (again, setting $\beta=\zeta=0$). These
simulations are performed from initial data near uniform isotropy. In
particular, the initial condition for $\Psi$ has the form
\begin{equation}
\Psi(x,\theta,0)= \frac{1}{2\pi}\left(1 +\underset{i}{\Sigma}\epsilon_{i}
\cos(\mathbf{k}_{i}\cdot\mathbf{x} +\varphi_{i})P_i(\theta)\right),
\label{eq:ic}
\end{equation}
where the amplitudes $\epsilon_{i} \in [-0.01;0.01]$ and phases
$\varphi_i \in [0, 2 \pi)$ are sampled from uniform distributions,
  with $P_i$ a low-order trigonometric polynomial which satisfies
  $\Psi\left(\theta+\pi\right)=\Psi\left(\theta\right)$.  We include
  only the 10 longest-wavelength spatial modes in
  Eq.~\eqref{eq:ic}. The initial data for the $BQ$ system is simply
  found by forming $\D$ from $\Psi(\x,\theta,0)$.

In agreement with analytical prediction, we find that the initial
concentration fluctuations decay, and the concentration becomes
uniform.  Figure~\ref{fig:evolution}(a) and (c) show the nematic
director field $\m(\x,t)$ overlaying the scalar order parameter
$s(\x,t)$. The system evolves quickly away from
the initially isotropic state to a temporally fluctuating state with
high local order.  Panels (b) and (d) show the fluid velocity field
$\u=(u,v)$ overlaying the planar (scalar) vorticity
($\omega=v_x-u_y$). The temporal and spatial dynamics for both models
are complex and seemingly chaotic, and qualitatively similar to those
seen in models of bacterial Pusher suspensions
\cite{aranson07,SS2008b} and of biofilament/molecular-motor assemblies
\cite{TGY2013,giomi13,GBGBS2015a,GBGBS2015b}.

While both the kinetic and $BQ$-tensor models start with the same data
for $\D$, the system dynamics being chaotic means that the two
descriptions diverge in their details. However, both descriptions do
evolve towards a similar statistical structure, as shown by the time
evolution in the mean nematic order $S(t)=\int_\Omega s(\x,t) dA/{\rm
  area}(\Omega)$, from near zero to a temporally fluctuating state
with mean near 0.6 (note the $BQ$ theory slightly overestimating the
order). The linear theory in Section \ref{lineartheory} predicts that
the kinetic and $BQ$-tensor models have the identical linear growth rates from a uniform isotropic base state. Exponential growth to
saturation is seen in $S(t)$ following a short equilibration period
where high wave-number modes rapidly decay.  Fitting $S(t)$ during the
growth period yields a growth rate $\sigma\approx0.16$ for both
models. This is close to the growth rate of the first fundamental
mode in the simulation box, $\sigma\approx0.167$, predicted by the
linear theory. This figure also suggests that the growth rate of the
first mode sets the time scale to saturation.

That the kinetic and $BQ$-tensor theories produce the same magnitude
of flow is seen in the very similar scales of their vorticity
dynamics. To quantify the flows' length scales, the inset to
Fig. \ref{fig:lines} plots the normalized space- and time-averaged
velocity-velocity spatial autocorrelation function
${\rm{Cor}}[\u,\u](R)=\left\langle\u(\x)\cdot\u(\x+\R)/|\u(\x)|^2\right\rangle$
where $R=|\R|$.  We see a close match in this measure between the
two models, as well as a slight negative minimum (indicating
oppositely directed flow at $R\approx 13$, which is near the box
size). This is consistent with the first fundamental spatial mode in
the box growing the most rapidly.

In comparing Figs.~\ref{fig:evolution}(a) and (c), we find in both models
large regions of high nematic order (light to dark red) and director
alignment. In the corresponding velocity fields ((b) and (d)),
these regions are associated with likewise oriented extensional
straining flows (which are necessarily of small vorticity). As found
in simulations of active nematics \cite{GBGBS2015a,GBGBS2015b}, both
models show $\pm 1/2$-order disclination defects in the $\m$ field,
inhabiting regions of small nematic order, and co-located with fluid
jets between oppositely signed vortical regions.

In Fig.~\ref{fig:evolution_dense}, we show a late-time comparison
for Model II using the kinetic theory and its $BQ$-tensor closure. Panels
(a) and (c) show clearly that, in both models, the nematic field contains
motile disclination defect pairs of order $\pm 1/2$. The
induced active flows in panels (b) and (d) look similar to those of
Model I, but with stronger vorticity.  The defects shown here exist in
regions of small nematic order (dark blue), and are born as opposing
pairs in elongated ``incipient crack'' regions
\cite{GBGBS2015a,GBGBS2015b}.  These crack structures locally decrease
nematic order, and increase curvature of director field
lines. Characteristically, we find that the $+1/2$ defects propagate
approximately along their central axis and have a much higher velocity
than those of $-1/2$ order. The relatively higher flow velocity in the
neighborhood of a $+1/2$ defect appears as a well-localized jet, in
the direction of defect motion, between two oppositely signed
vortices. These dynamics are similar to those observed previously in
studies of polar kinetic models of microtuble/motor-protein
suspensions \cite{GBGBS2015a,GBGBS2015b}, and other active nematic
models \cite{TGY2013,giomi13,Giomi2015}.

Again, as shown by the evolution of the mean nematic order parameter
in Fig. \ref{fig:lines}, the kinetic and $BQ$-tensor models, evolving
from the same $\D$ data, show statistically similar long-time
dynamics, and nearly identical growth from isotropy. As expected, we
observe a more rapid growth in Model II away from isotropy due to the
Maier-Saupe alignment torques; see Eq.~(\ref{growth rates}). The
kinetic and $BQ$-tensor models again show nearly the same exponential
growth to saturation of $S$, with a fit yielding $\sigma\approx 1.2$, close to the linear growth rate of the first
fundamental mode in the box $\sigma\approx 1.08$ (which gives a
time scale for growth to saturation).  Finally, the velocity-velocity
autocorrelation function shows excellent agreement between the
kinetic model and its $BQ$-tensor closure (and improved over Model I).

\subsection{Active Flows in Confined Geometries}


Having established an excellent correspondence between the kinetic and
$BQ$-tensor theories, we now study the dynamical behavior of an active
suspension confined to an enclosure, using the $BQ$-tensor version of
Model II only.  Experiments have used motile suspensions of bacteria
\cite{Lushi2014}, Quincke rollers \cite{bricard13}, and
MT/motor-protein assemblies \cite{SanchezEtAl2012} to study the
effects of confinement. Previous work \cite{woodhouse2012,ezhilin2015}
has studied bacterial suspensions using models closely related to the
$BQ$-tensor model of Model I. To establish correspondence with these
works, in Appendix~\ref{AppendixC} we compare the dynamics of Model I
in a circular chamber to previous analytical and numerical results by
Goldstein and Woodhouse \cite{woodhouse2012}. Overall, we find
excellent agreement, showing a bifurcation with decreasing $\alpha<0$
from isotropy to an axisymmetric swirling state, further to a symmetry
breaking associated with +1/2-order disclination singularities, and
thence to complex aperiodic dynamics. Here, within the context of
Model II, we focus on the interaction of active extensile stresses and
steric ordering with the geometry of the confining chamber.

\begin{figure}
\begin{center}
  \includegraphics[width=1.0 \textwidth]{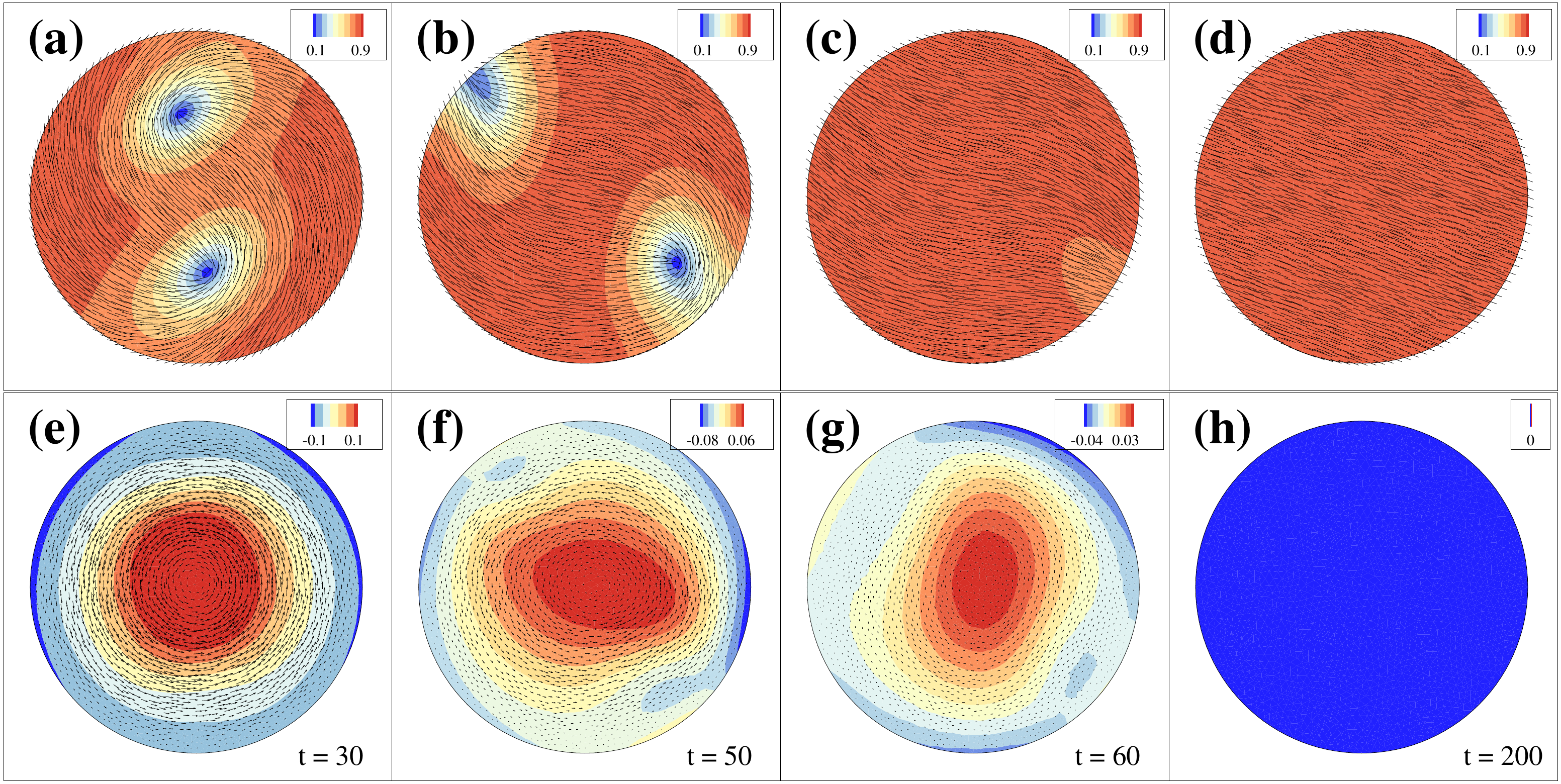}
\end{center}
\caption{Evolution of an Extensor suspension from near isotropy, for small but negative $\alpha$ ($\alpha=-0.3$),
  when confined to a disk of radius $R=2.0$. The upper row shows the
  nematic field $\m$ overlaying the scalar order parameter $s$ (color
  field), while the lower row shows the velocity vector field $\u$
  overlaying its scalar vorticity (color field).}
  \label{fig:concentrated1}
\end{figure}

\smallskip
\noindent{\bf Boundary Conditions.} To solve the confined equations,
we must supplement Eqs.~(\ref{Bingham1})-(\ref{Bingham3}) with
appropriate boundary conditions. We choose the simplest that conserve
total particle number. Consider a bounded 2D fluid domain $\Omega$
with boundary $\partial\Omega$; for example, see the disk in
Fig.~\ref{fig:concentrated1}. Here we assume the no-slip condition
$\u|_{\partial\Omega}=\0$ for the velocity field when solving the
Stokes equation. For the $\D$ advection-diffusion equation we return
to the kinetic description in
Eqs.~(\ref{eq:smol},\ref{xflux},\ref{pflux}), and find that
conservation of total particle number is enforced by the boundary
condition $(\u\Psi-d_T\nabla\Psi)|_{\partial\Omega}\cdot\n=0$, which
reduces to $d_T\nabla\Psi|_{\partial\Omega}\cdot\n=0$. Taking the
second moment with respect to $\p$ then gives the zero-flux boundary
condition $(\n\cdot\nabla)\D=\0$ on $\partial\Omega$. Note that this
condition does not enforce any particular alignment direction at the
boundary.

\smallskip
\noindent{\bf Numerical Method.} We solve this finite-domain problem
using a finite element method with an unstructured triangular grid.
The governing equations are discretized using the standard 2D Galerkin
formulation:
\begin{align}
  &\int_\Omega  {\left( {{\mathop{\rm Re}\nolimits}
\frac{{\partial {\bf{u}}}}{{\partial t}} \cdot {\bf{\tilde u}}
- p\nabla  \cdot {\bf{\tilde u}}
+ {\bsig[\D]} :\nabla {\bf{\tilde u}}} \right) } dA = 0\\
  &\int_\Omega  { \left( {\nabla  \cdot {\bf{u}}} \right)} \tilde p~ dA  = 0\\
  &\int_\Omega {\left( {{\D^\nabla } + 2\nabla \u:\S
- 4{\zeta}\left({\D \cdot \D - \D:\S} \right) - {d_T}{\nabla ^2}\D
+ 4d_R\left( {\D - \frac{\I}{2}} \right)} \right)} :{\bf{\tilde
      D}}d\Omega = 0
\end{align}
where $\tilde {\textbf{u}}$, $\tilde {p}$, and $\tilde \D$ are test
functions. Here we are solving an unsteady Stokes equation with a
partial derivative term $\rm{Re} \left(\frac {\partial \bf u}{\partial
  t}\right)$ with an estimated Reynolds number of $\rm{Re} =
10^{-3}$. In the finite element solver, we choose different
interpolation functions for different unknown variables in order to
satisfy the LBB condition \cite{Gao09,Hu01}. The fluid velocity is
approximated by piecewise quadratic functions which are continuous
over $\Omega$ (P2). The pressure and stress are approximated by
piecewise linear functions (P1). For this type of mixed finite
elements, it is known that when $\rm{Re}$ is small, the numerical
solution of the flow field can achieve third order accuracy in space
\cite{Thomasset81}. For this study, the temporal discretization is a
second-order difference scheme.  The governing equations are reduced
to a nonlinear system of algebraic equations which is solved by a
modified Newton-Raphson algorithm \cite{Hu01,Gao09}.  Specifically, in
each iteration of the algorithm, the discretized linear system is
solved by the GMRES method \cite{Saad86}. An incomplete LU
preconditioner \cite{Chow97} is used to accelerate convergence. In the
following, we mainly vary parameters $\alpha$ and $R$ but fix $\zeta =
0.5$, $\beta = 0.874$, and $d_T=d_R=0.05$.

\begin{figure}
\begin{center}
  \includegraphics[width=1.0 \textwidth]{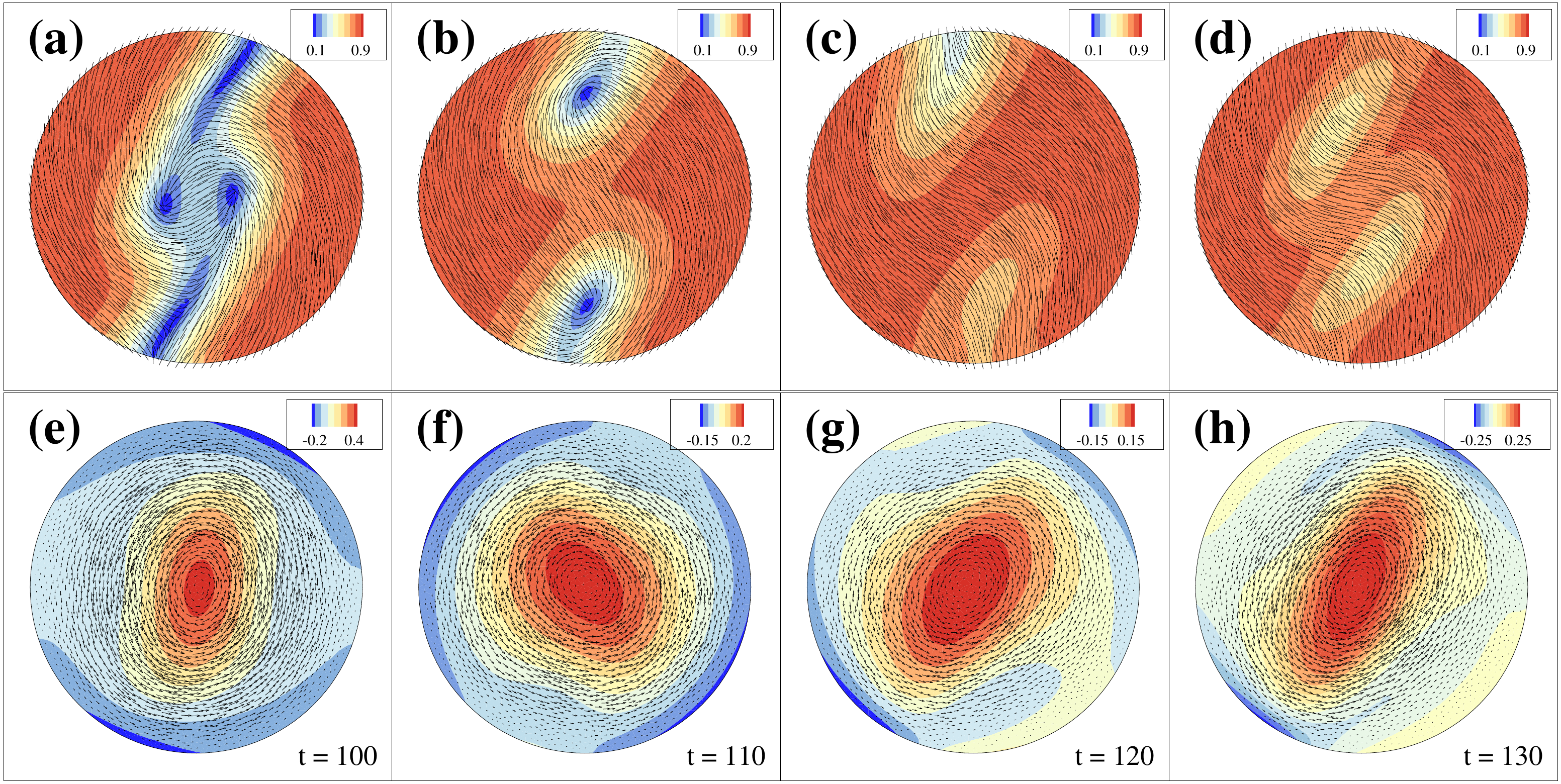}
\end{center}
\caption{Evolution of an Extensor suspension from near isotropy with $\alpha=-0.7$, confined to a disk of radius
  $R=2.0$. The upper row shows the nematic field $\m$ overlaying the
  scalar order parameter $s$ (color field), while the lower row shows
  the velocity vector field $\u$ overlaying its scalar vorticity
  (color field); see supplementary video S1.}
  \label{fig:concentrated2}
\end{figure}

\begin{figure}
\begin{center}
  \includegraphics[width=1.0 \textwidth]{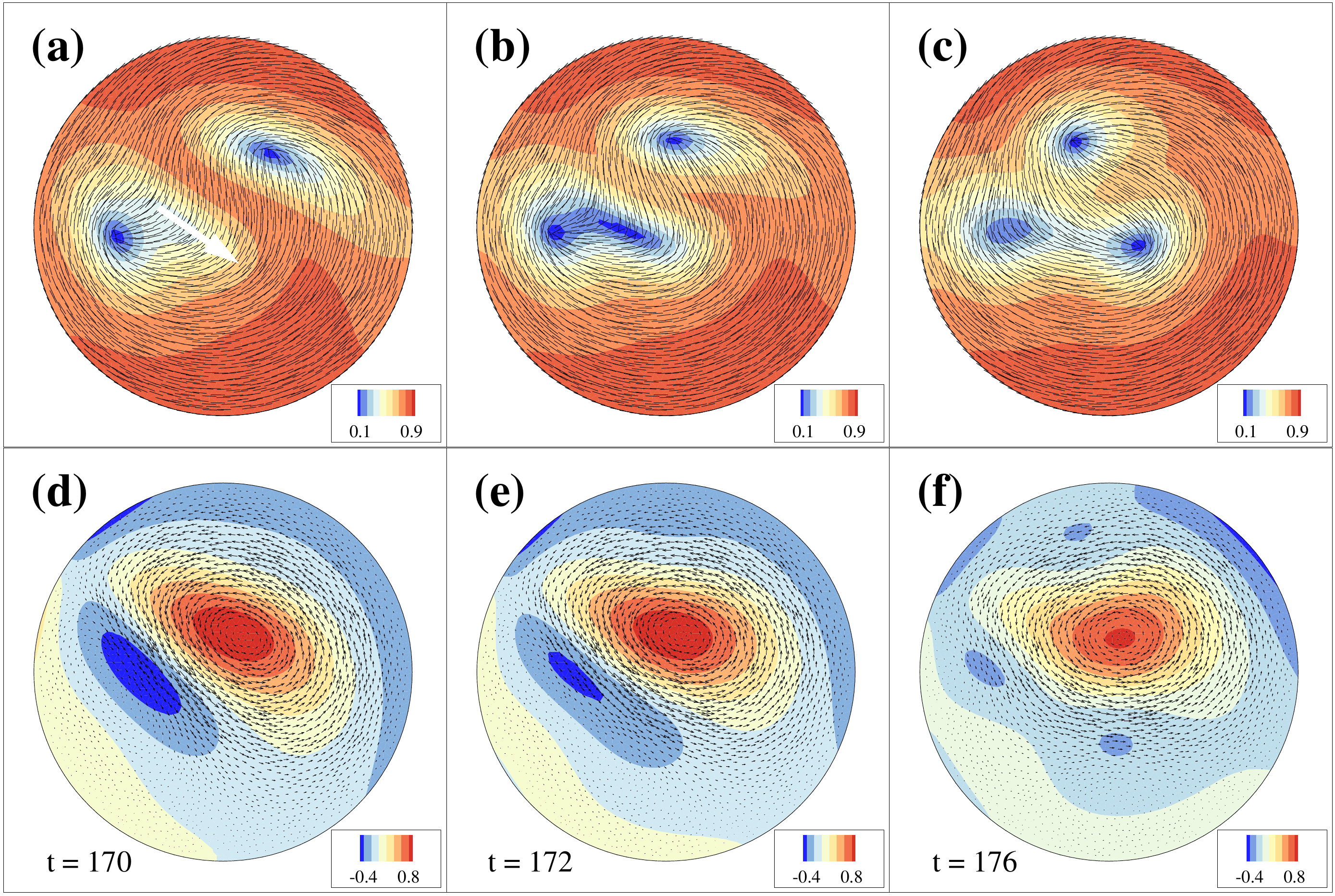}
\end{center}
\caption{Evolution of an Extensor suspension from near isotropy with $\alpha=-2.0$, confined to a disk of radius
  $R=2.0$. The upper row shows the nematic field $\m$ overlaying the
  scalar order parameter $s$ (color field), while the lower row shows
  the velocity vector field $\u$ overlaying its scalar vorticity
  (color field); see supplementary video S2.}
  \label{fig:concentrated3}
\end{figure}

\begin{figure}
\begin{center}
  \includegraphics[width=0.45 \textwidth]{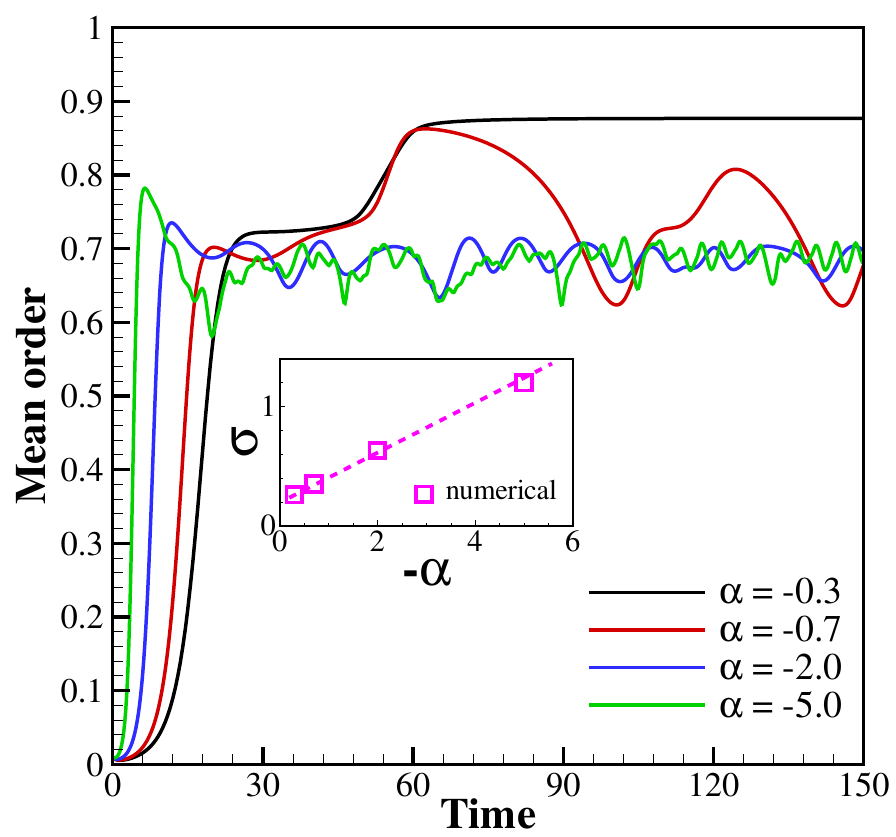}
\end{center}
\caption{Evolution of the mean nematic order $S(t)$ at different
  levels of activity with $R=2$, corresponding to the cases shown in
  Figs.~\ref{fig:concentrated1}-\ref{fig:concentrated3}, and
  supplementary videos S1-S3. The initial growth of $S$ is found to be
  exponential in time, and the inset shows a fit to the growth rates
  $\sigma$ (open squares) as a function of $-\alpha$. The dashed
  linear fit suggests that the growth rate scales approximately with
  $|\alpha|$.}
\label{fig:meanorder_dense}
\end{figure}

\begin{figure}
\begin{center}
  \includegraphics[width=1.0 \textwidth]{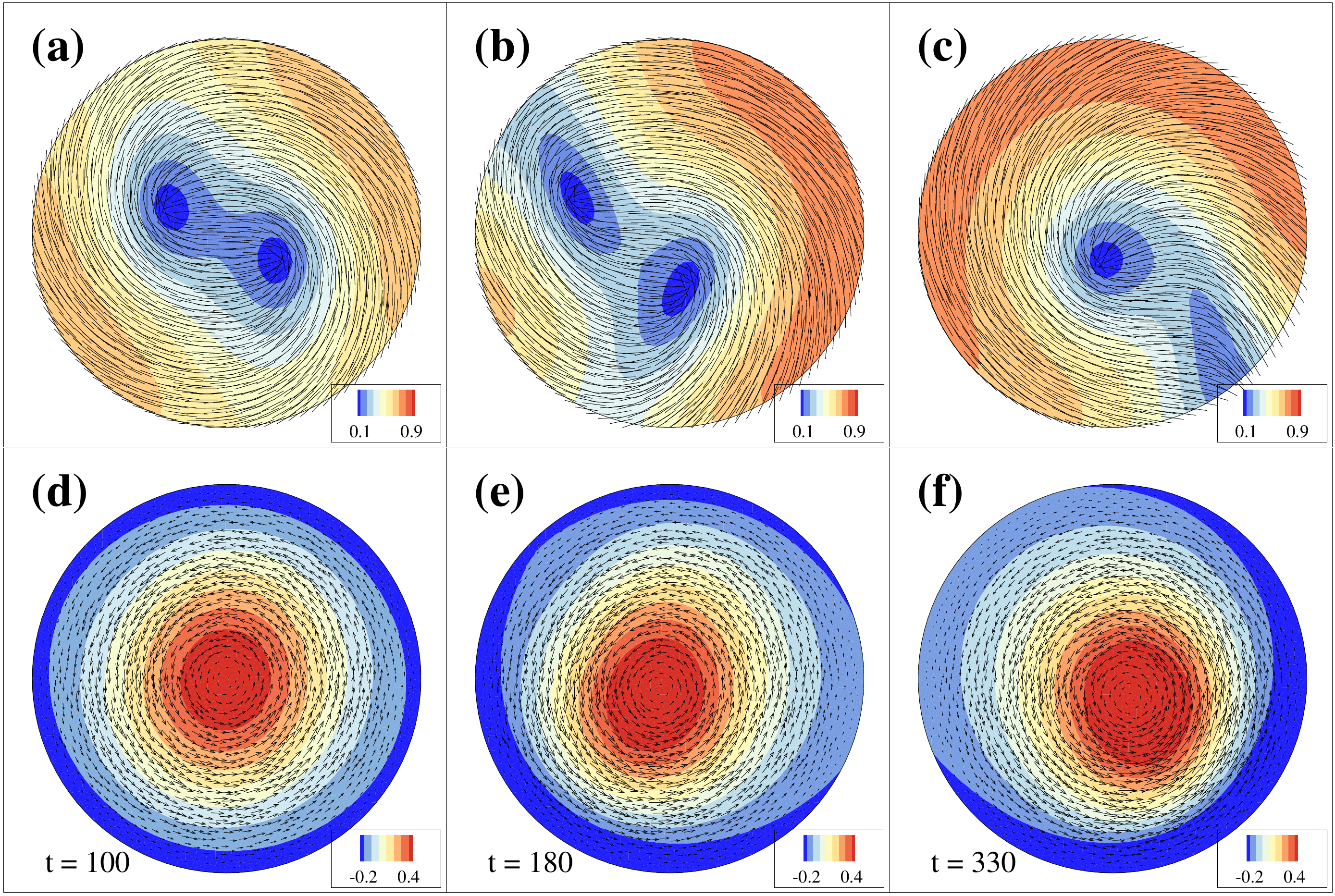}
\end{center}
\caption{Evolution of an Extensor suspension from near isotropy with $\alpha=-2.0$, when confined to a smaller disk of radius $R=1.25$. The upper row shows the nematic field $\m$ overlaying the
  scalar order parameter $s$ (color field), while the lower row shows
  the velocity vector field $\u$ overlaying its scalar vorticity
  (color field); see supplementary video S4.}
  \label{fig:concentrated4}
\end{figure}

\subsubsection{Active flows in a circular chamber}

To start, in Fig.~\ref{fig:concentrated1} we consider an Extensor
suspension with weak activity, $\alpha=-0.3$. From a nearly isotropic
state, the orientation field rapidly orders (note the increase in the
scalar order parameter as compared to the cases in Appendix
\ref{AppendixC} where $\zeta = \beta =0$ are used in Model I) and two
$+1/2$-order disclination defects appear (panels(a,e)), moving about
the center (as seen in Fig.~\ref{fig:dilutedfct}b), while a vortical
field appears concomitantly. However, these two defects spiral
outwards to the disk boundary where they are absorbed (panels
(b,f)). No new defects are thereafter created and the vortical flow
weakens (panels (c,g)) as the nematic orientation field relaxes to a
homogeneously aligned state (panels (d,h)), which is allowed by the
zero-flux boundary condition for $\D$.

Increasing activity (i.e., decreasing $\alpha$) leads to persistent
dynamics, as shown in the accompanying supplementary videos. In Fig.~\ref{fig:concentrated2},
$\alpha$ has been decreased to -0.7; see video S1 in the supplementary
information. Initially it follows the course of the previous case: the
system moves quickly away from isotropy with the appearance of two
$+1/2$-order disclination defects and a vortical flow. The defects
spiral outwards, are absorbed by the walls, and the system relaxes
into a state of nearly homogeneous nematic alignment. However, this
state is unstable to hydrodynamic instabilities driven by activity
\cite{GBGBS2015a}. As seen in video S1, two cracks of low nematic order
appear, with each birthing a pair of disclination defects of opposing
sign (order $\pm 1/2$; panels (a,e)). The $-1/2$-order defects are now
absorbed by the wall (panels (b,f)), followed shortly thereafter by
the $+1/2$-order defects (panels (c,g). The process begins anew with
the bending of nematic field lines and the formation of two low nematic order
cracks (panels (d,h)). Accompanying the nematic structure variation,
we observe the persistence of background rotational flows which
strengthen and weaken through the various stages of this periodic
dynamics (lower row of panels).

The dynamics becomes quasi-periodic or chaotic at yet higher
activity. For $\alpha=-2$, Fig.~\ref{fig:concentrated3} shows a short
period in time that illustrates the
complex system dynamics (see also supplementary video S2). In
panels(a,d) we see that as $\alpha$ further decreases, there is still
a basic dynamics of two rotating $+1/2$-order defects. The rotational
flow is more finely scaled as there is now a secondary vortex of the
opposite sign to the primary (red) counter-clockwise vortex. The
dipolar form of the vortical distribution is associated with a strong
fluid jet, marked as the white arrow in panel (a). The primary vortex
is more compact, leading to a somewhat smaller circulatory flow.
Shortly thereafter (panels (b,e)), a low-order crack appears near one
of the defects and births a defect pair. The newly formed $-1/2$-order
defect merges with and annihilates the older $+1/2$-order defect
nearby (or sometimes both defects are absorbed into the
wall), and the process begins anew; see panels (c,f).

Finally, we refer the reader to the supplementary video S3 for
$\alpha=-5$, which shows the yet more complex interplays between
defects, cracks, and the defect pairs thereby produced. The associated
vortical flows are no longer dominated by a single-signed vortex but
is much more dipolar, with a fluctuating predominance of one
signed vortex over the other. While complicated both spatially
and temporally, the dynamics nonetheless maintains a
quasi-periodicity.

Figure \ref{fig:meanorder_dense} shows the evolution of the mean
nematic order parameter $S(t)$ for these various simulations. In all
cases the rapid growth of $S(t)$ away from zero reflects the
transition from isotropy to a state of higher nematic order. For the
simulation with the lowest activity, $\alpha=-0.3$ (black curve), this
initial rise yields a transitory plateau slightly above $S=0.7$ during
which the dynamics is characterized by formation and outward
spiraling of two +1/2-order defects. Their absorption by the chamber
walls, when $40<t<60$ (see Fig.~\ref{fig:concentrated1}), leads to
relaxation to a homogeneously ordered state with $S$ slightly less
than 0.9. The emergence of persistent and complex defect dynamics in
the other cases is reflected in the saturated behaviors of $S(t)$
following growth from isotropy. In all cases the initial growth from
zero can be well-fitted by an exponential in time. The calculated
growth rates are plotted in the inset, and show a linear
increase with increasing activity $\alpha$.

We have also examined the dynamics in circular chambers of different
radii and found generally similar behaviors. That being said, in a
smaller confinement we have also discovered a fascinating periodic
coherent structure. Figure~\ref{fig:concentrated4} shows the dynamics
of the system for $\alpha=-2$ ({\it cf.} Fig.~\ref{fig:concentrated3})
but with confinement radius $R=1.25$, as it enters into a periodic
dynamics. The defect dynamics starts from the apparent splitting of a
$+1$-order defect into two $+1/2$ defects (panel (a); see
supplementary video S4) co-rotating with the associated nearly
axisymmetric vortical field (panel (d)). The two defects then
destabilize from a common rotation centered about the origin (panels (b,e)), and begin to, apparently, move into and out of the wall (i.e., the wall both absorbs and produces defects). The apparent production of
defects at confinement boundaries has been observed in experiments
involving microtubule/motor-protein assemblies (private communication with
Z. Dogic). Eventually, the system relaxes to having one defect
seemingly trapped outside of the physical flow domain, leaving only a
vestigial low-order crack which moves clockwise around the boundary in
the direction opposite to the counter-clockwise circulation of the flow
(panels (c,f)).

\begin{figure}
\begin{center}
  \includegraphics[width=0.7 \textwidth]{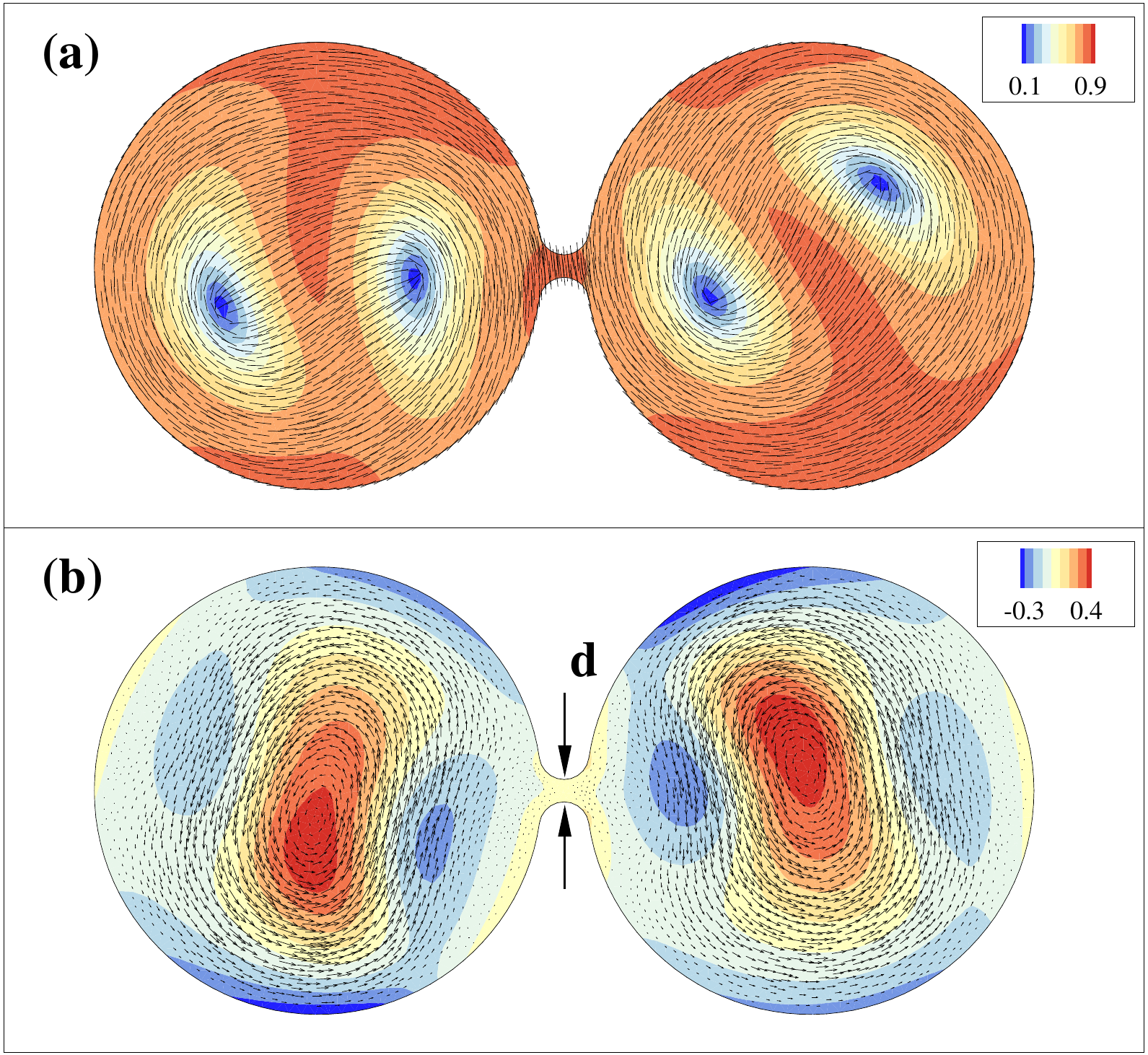}
\end{center}
\caption{For two circular chambers connected by a narrow neck of width
  $d=0.2$, and filled with an Extensor suspension with $\alpha=-2$, (a)
  the nematic vector field $\m$ overlaying the scalar order parameter
  $s$ (color field), and (b) the fluid velocity field $\u$ overlaying
  its scalar vorticity $\omega$. See supplementary video S5.}
  \label{fig:bicon1}
\end{figure}

\begin{figure}
\begin{center}
  \includegraphics[width=0.7 \textwidth]{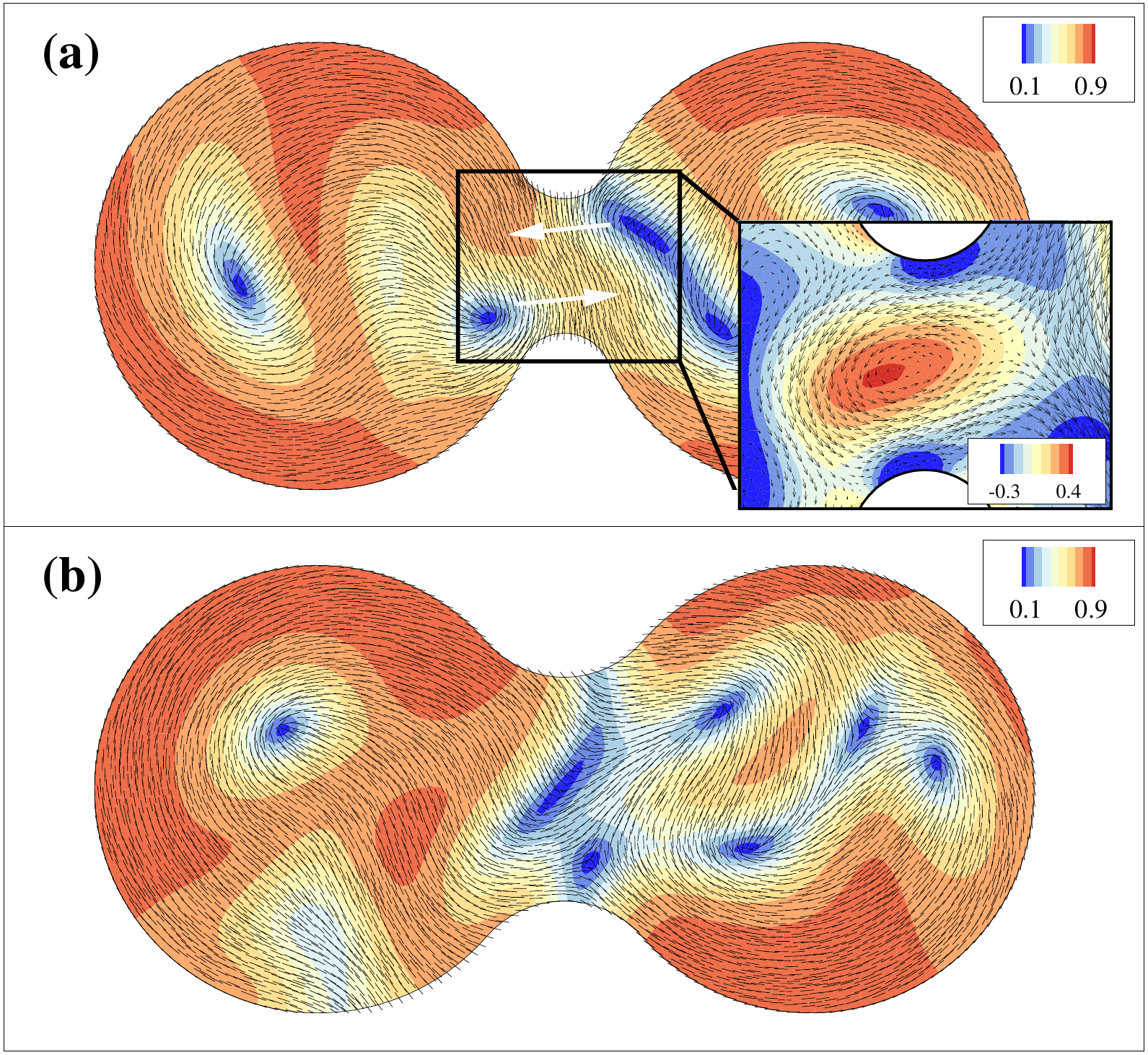}
\end{center}
\caption{Two circular chambers connected by a neck of width $d$, and
  filled with an Extensor suspension with $\alpha=-2$. (a) For $d=1.2$,
  the nematic vector field $\m$ overlaying the scalar order parameter
  $s$ (color field).  The arrows show the direction of shearing of the
  nematic field lines. The inset shows the fluid velocity field $\u$
  overlaying its scalar vorticity $\omega$ in the neck region. The
  arrows also correspond to the location and direction of fluid jets
  between the two chambers. See supplementary video S7. (b) The
  nematic vector field and scalar order parameter for $d=2.0$. See
  supplementary video S8.}
  \label{fig:bicon2}
\end{figure}

\subsubsection{Active flows in a biconcave chamber}

The coarse-grained $BQ$-tensor model and its finite-element
discretization greatly facilitates the exploration of active nematic
flows in more complex geometries. In Fig.~\ref{fig:bicon1}, we study
the dynamics of an Extensor suspension confined to a biconcave chamber
where two circular chambers, each or radius $R=2$, are connected
smoothly by a bridge of width $d$.  Here we fix $\alpha=-2$. Supplementary videos S5-S8 show the dynamics as $d$ is
successively increased ($d=$0.2, 0.8, 1.2, and 2). We find that when
the neck is thin, the dynamics in the two chambers are very similar to
the case shown in Fig.~\ref{fig:concentrated2} for $\alpha=-2$, and
appear to be evolving independently of each other; see supplementary
video S5.  This lack of interchamber communication is consistent with
the velocity stagnation zone in the bridge (panel (b)). While a thin
neck would be sufficient to prevent interplay between the two
chambers, this effect is likely reinforced by the alignment of the
nematic field lines across the neck which lends elastic rigidity to
the material contained therein.

The essentially independent dynamics of the two chambers persists to
quite wide necks, e.g. $d=0.8$. As seen in supplementary video S6, the
nematic field lines still span the neck, apparently blocking any
substantial breaching flows. As $d$ is further increased, beyond
$d_{crit} \approx 1.0$, the dynamics in the two chambers begins to
couple. Figure~\ref{fig:concentrated2}(a) and supplementary video S7
for $d=1.2$ show significant bending and shearing of the nematic field
lines, with the panel(a) inset showing that these deformations are
associated with a vortex bounded above and below by two fluid jets
that penetrate from one side to the other. For the largest neck width,
$d=2.0$, Figure~\ref{fig:concentrated2}(b) and supplementary video S8
show that flows now move freely between the two chambers.
This transition is reminiscent of a Fredricks transition where an
electric or magnetic field needs to exceed a critical intensity to
deform a uniformly aligned nematic liquid crystal material spanning
between two plates. Fredericks transitions have been studied in other
contexts in the field of active matter \cite{giomi08,Voituriez2005}.

\section{Discussion}
\label{sec:conclusion}

We explored a multi-scale model, and its
simplifications, for suspensions of rod-like particles that exert
active dipolar extensile stresses on the immersing solvent. This model
is relevant to suspensions or bundles of microtubules that undergo
polarity sorting driven by crosslinking motor proteins
\cite{surrey01,bendix08,hentrich10,gordon12}, and directly models
suspensions of chemically-active particles whose consumption of a
chemical fuel creates extensile flows along the particle
\cite{Wykes16,Jewell2016,Pandey2016}. Aside from analysis of the
model, a central goal here was to investigate the Bingham closure
\cite{Bingham74,Chaubal98,Feng98} as applied to active suspensions. We
found that the resultant $BQ$-tensor theory gave an excellent
accounting of the chaotic self-driven flows of active suspensions
(both with and without strong alignment interactions), and allowed us
to investigate with relative ease the behavior of an active-nematic
fluid under confinement.  We found several fascinating
behaviors, such as defects being absorbed at (and perhaps birthed by)
boundaries, complex rotational flows, and an apparent Fredericks
transition associated with nematic elasticity.

We remark that the $BQ$-tensor theory has no unfixed constants with
respect to the full kinetic theory, and is far cheaper to simulate
(essentially by a factor inversely proportional to the number of
points of the sphere (3D) or circle (2D) used to resolve the
distribution function in orientation). This also establishes a
connection between Doi-Onsager kinetic theories and $Q$-tensor
theories such as those based on the Landau-deGennes approach
(e.g. \cite{giomi13,Giomi2015,LWZ2015}). We are currently working on improving such
closure approaches, from a variational perspective, and extending them
to polar kinetic theories such as for motile suspensions
\cite{SS2008a,subramanian09} and microtubule/motor-protein suspensions
\cite{GBGBS2015a,GBGBS2015b}.

\subsection{Length scale determination}

\begin{figure}[htp]
  \makebox[\textwidth][c]{\includegraphics[width=0.45 \textwidth]{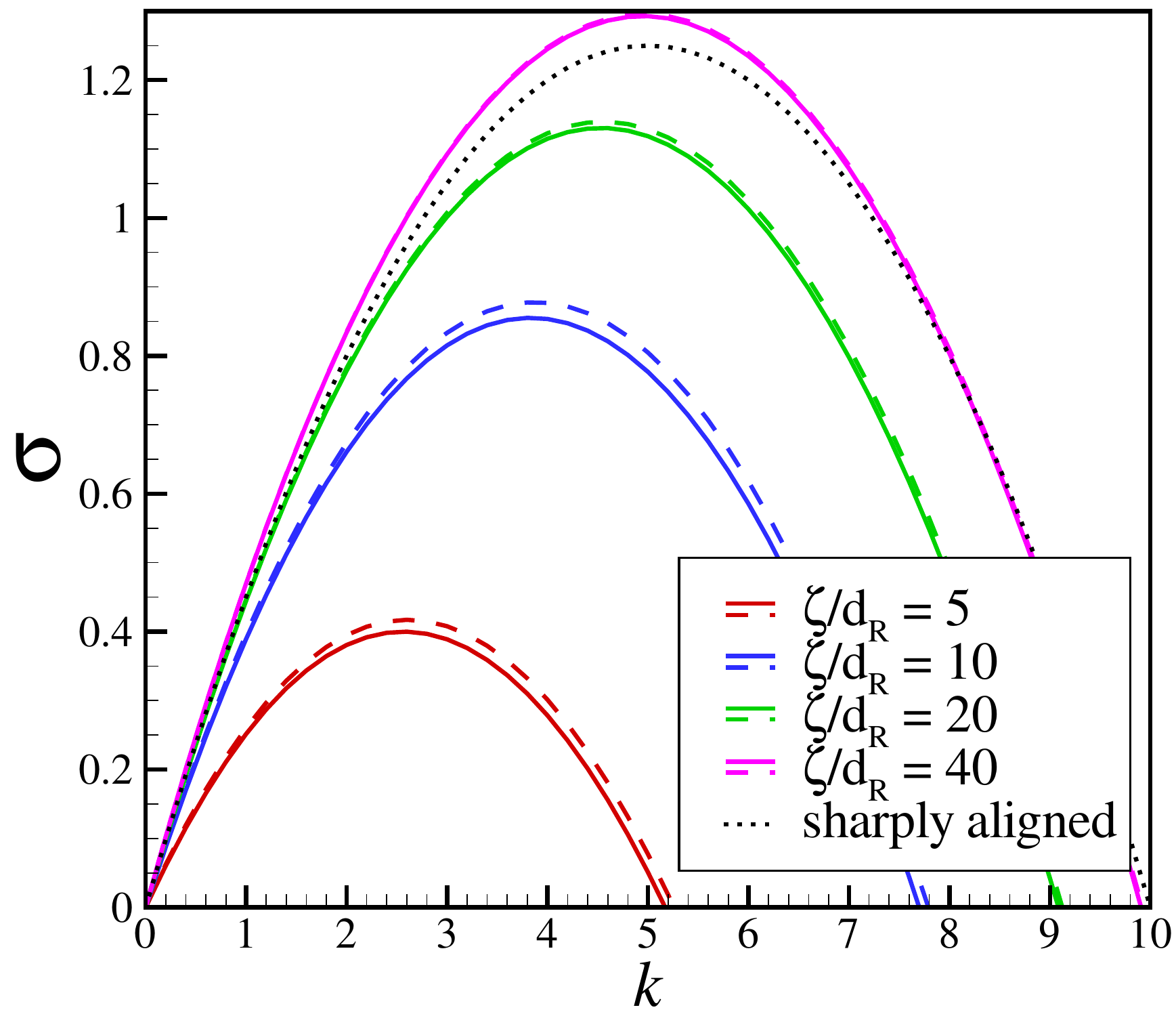}}%
  \centering
  \caption{Linear stability analysis for a thin layer of homogenous
    nematically-aligned Extensor suspension confined between two
    fluids. growth rates are plotted as a function of wave-number $k$
    at various $\zeta/d_R$ using both the $BQ$-tensor theory (dashed
    lines) and the kinetic theory (solid lines). The black
  dotted curve is the growth rate predicted from the sharply-aligned case. Computation
    parameters are chosen as $\alpha=-1.0$, $\beta = 0.874$, and
    $d_T=d_R=0.05$.  }
  \label{fig:surface}
\end{figure}

We close with a short digression into the determination of
length scales of instability in active suspensions, as a suspension of
Extensors is a particularly simple case to discuss. As previously
discussed, it seems generic that bulk models of active suspensions
lack an intrinsic length scale determined by a most unstable mode
\cite{simha02,SS2008a,GBGBS2015a}.  This is seen most directly in
Eq.~(\ref{growth rates}) for the growth rates of plane-wave
perturbations from isotropy, and in our stability analyses of aligned
states. It is also consistent with our simulations in periodic domains
where observed growth rates from isotropy are well-described by that
for the first fundamental mode with $k=2\pi/L$ of the periodic
domain. Nonetheless, some studies have sought to identify an intrinsic
length scale of instability \cite{thampi14b}. Such a scale is present when
the system is subject to an external drag or damping, as when an
active material is confined to a thin layer between two fluids (e.g.,
Gao {\it et al.}  \cite{GBGBS2015a} in modeling the experiments of
Sanchez {\it et al.} \cite{SanchezEtAl2012}, as well as Leoni and
Liverpool \cite{leoni10}), or sits atop a substrate
\cite{Thampi2014c}.

To be specific, consider a thin active layer evolving in the
$xy$-plane at $z=0$, across which layer activity gives a
tangential stress jump between two outer 3D Stokesian fluids. By
taking a 2D $xy$ Fourier transform of the 3D Stokes equations, the two
half-space problems for the 3D Stokes flows can be solved analytically
under the conditions of zero velocity as $|z|\rightarrow\infty$ and
continuity of velocity at $z=0$. This determines the interfacial
velocity field $\u(x,y)$ as a function of the active layer
extra stress $\bsig$ \cite{MS2014,GBGBS2015a,GBGBS2015b}:
\begin{equation}
\tilde{\u}(\k) = \frac{i}{2k}\left(\I-\hat{\k}\hat{\k}\right) \cdot
  \tilde{\bsig}  \cdot \k
\label{eq:surface}
\end{equation}
where ${\bf{\hat k}} = {{\bf{k}}}/{{\left| {\bf{k}} \right|}}$ is the
2D unit wavevector. While not strictly necessary, we have assumed that the interfacial flow is incompressible (as seems consistent with
the experiments of Sanchez {\it et al.} \cite{SanchezEtAl2012},
private communication with Z. Dogic). Thus, activity at the interface
drives flows in the two outer fluids, which in turn provides a drag on
the active surface. Simulating this situation amounts to replacing the
inversion of the 2D Stokes operator in Eq.~(\ref{eq:regular}) with
Eq.~(\ref{eq:surface}).

The inversion formula in Eq. \eqref{eq:surface} differs by a factor of
$k/2$ from that in Eq. \eqref{eq:regular} where the Stokes equation is
forced by a bulk stress.  The propagation of that difference through
the stability analysis is found by examining the coefficients $\gamma$
and $\omega_k$ (Eq.~(\ref{LinearD1})) that govern the growth/decay
rate $\sigma_k=-(\gamma+\omega_k)$ (Eq.~(\ref{LinearD2})). For the 2D
(or 3D) bulk fluid case, $\gamma$ has the form
$\gamma=\tilde{\alpha}/(1+\tilde{\beta})$; while for the immersed
interface case considered here, it is modified to
$\gamma_k=(\tilde{\alpha}k/2)/(1+(\tilde{\beta}k/2))$ which rises
monotonically in $k$ from zero to $\tilde{\alpha}/\tilde{\beta}$.  For
${\tilde\alpha}<0$, competition of activity with translational diffusion
in $\omega_k$ yields a unique critical wavenumber $k_{c}>0$ at which
$\sigma_k$ is maximized.

This same effect manifests itself in the instability of a homogenous
nematically-aligned state, and yields an intrinsic scale that is
observed in nonlinear simulations (see Gao {\it et al.}
\cite{GBGBS2015a} for the more complicated case of
microtubule/motor-protein suspensions). For a simple Extensor
suspension, Fig.~\ref{fig:surface} shows the numerically calculated
growth rates $\sigma_k$ (red curve) of a plane-wave perturbation to
the homogeneous nematic state, with wave-vector taken in the direction
of maximal growth, $\theta=0$.  This being the direction of maximal
growth is in agreement with the sharply-aligned analysis of
Sect.~\ref{Sharp}, and has been found in other studies of active
suspensions \cite{SS2008a,SS2008b,ESS2013,GBGBS2015a,GBGBS2015b}. The
computed curve also shows an intrinsic wave-number of maximal growth,
as was found analytically for the isotropic state.  As $\zeta$ is
increased for fixed $\alpha$, we see an increase in maximal
growth rate and decrease in its associated length scale.  In that limit we find an increasing correspondence between the full
kinetic theory and the $BQ$-theory.

The parabolic-like structure of the growth rate curve is in
qualitative agreement with our phenomenological sharply-aligned model
when modified to the immersed interface case. This modification alters
the plane-wave growth rates $\tilde{\sigma}_{1,2}$ in
Eqs.~(\ref{ModifiedLinearGrowthAligned})(where $\beta=0$) to the form
\begin{equation}
  \hat{\sigma}_{1,2}=-\alpha H_{1,2}(\Phi)k/2-d_Tk^2.
  \label{ModifiedLinearGrowthAligned2}
\end{equation}
For $\alpha<0$, this expression yields at each $\Phi$ (with $H$
maximal at $\Phi=0$) a maximal growth rate
$\sigma_{max}=\alpha^2H^2(\Phi)/16d_T$, occurring at $k_{c}=-\alpha
H(\Phi)/4d_T$. To emphasize an important point: the
quadratic increase in maximal growth rate with activity follows from $k_c$
increasing linearly with activity (i.e., decreasing negative
$\alpha$). For the bulk case, where the system size sets the most
unstable scale, maximal growth rates should scale linearly with
activity (as in Eqs.~(\ref{growth rates}),
(\ref{ModifiedLinearGrowthAligned}), and the inset to
Fig.~\ref{fig:meanorder_dense}).

{\it Acknowledgements.} This work was funded by NSF grants DMR-0820341
(NYU MRSEC: TG, MS), DMS-1463962 and DMS-1620331 (MS), DMS 1619960 (TG, MDB, MS), DMR-0847685 (MDB), DMR-1551095 (MDB), DOE grant DE-FG02-88ER25053 (TG, MS), NIH grant K25GM110486 (MDB), R01 GM104976 (MDB, MS).

\appendix

\section{Stability of 3D uniform isotropic suspension}
\label{AppendixB}
\label{lineartheory}
While the linear theory of the Extensor suspension is a special case of that for a motile
one, it does have an especially simple and evocative structure that is
obscured in the motile case. For an isotropic steady state we have
$\Psi^0=1/4\pi$, $\D^0=\I/3$, $\u^0=\0$, and
$S^0_{ijkl}=\frac{1}{15}\left(\delta_{ik}\delta_{jl}+\delta_{il}\delta_{jk}+\delta_{ij}\delta_{kl}\right)$. By
perturbing this steady state as $\D=\I/3+\varepsilon
{\bf{D'}},\;{\bf{S}} = {{\bf{S}}^0} + \varepsilon {\bf{S'}},\;{\bf{u}}
= \varepsilon{\bf{u'}}$, the 3D linearized
Smoluchowski equation (\ref{eq:smol}) becomes
\cite{SS2008a,HS2010,ESS2013}
\begin{equation}
\label{Appeq:linSmol}
\frac{\partial \Psi'}{\partial  t}-\frac{3}{4 \pi} \p\p:\E'
=\frac{3\zeta}{2 \pi} \p\p:\D' + d_T \Delta \Psi'
+ d_R\nabla^2_{p}  \Psi'.
\end{equation}
This result uses that, for any $\p$-independent tensor $\A$,
$\nabla_p\cdot\left( (\I-\p\p)\cdot \A \cdot \p\right )=\mbox{tr}(\A)-3\p\p:\A$,
and that $\mbox{tr}(\D')=0$ for a constant density flow.
Multiplying Eq.~(\ref{Appeq:linSmol}) by $\p\p$, and integrating
over the unit $\p$-sphere gives the evolution equation for $\D'$:
\begin{equation}
\frac{{\partial {\bf{D'}}}}{{\partial t}} -\frac{2}{5} \E'
    = \frac{4\zeta}{5}{\bf{D'}}+ {d_T}{\Delta}{\bf{D'}}
    - 6d_R{\bf{D'}}.
\label{Appeq:Dprime}
\end{equation}
This result uses that $\S_0:\A=2\A/15$ for any symmetric trace-free
tensor $\A$. An identical result is obtained through linearization of
Eq.~(\ref{eq:qtensor}). The momentum balance equation and incompressible condition,
Eq. (\ref{eq:stokes2})-(\ref{eq:stress2}), linearize to
\begin{eqnarray}
\label{AppEq:linconcstokes}
\nabla \Pi-\left(  1+\frac{\beta}{15}\right)  \Delta\u'
&=&\left( \alpha-\frac{2}{5}\beta\zeta\right) \nabla \cdot \D', \\
\label{AppEq:linconcstokes_2}
\nabla\cdot\u'&=&0.
\end{eqnarray}

Rewriting Eq.~(\ref{AppEq:linconcstokes}) in terms of the stream
function allows significant simplification. By assuming that the
fluid domain is simply connected, we can define the vector stream
function $\bPhi$ such that $\u = \curl \bPhi$ and $\Delta \bPhi = -
\curl \curl \bPhi = -\bomega$. Taking the curl of
Eq.~\eqref{AppEq:linconcstokes} then yields
 \begin{equation}
\label{Appeq:concvecstream}
\nabla^{4}\bPhi=\tilde{\alpha} \nabla\times \nabla\cdot \D',
\end{equation}
where $\tilde{\alpha}=(\alpha-\frac{2}{5}\beta\zeta)/(1+\beta/15)$. Through
the definition of $\D'$ we can express the right-hand-side of this
equation, component-wise, in terms of $\Psi'$: \beq (\nabla\times
\nabla\cdot \D')_i = \int_S dS_{p}\ \epsilon_{ijk} \frac{\partial
}{\partial x_{j}}\frac{\partial }{\partial x_{l}} \left(p_k
  p_l\right)\Psi' = \int_S dS_{p}\ L_i \Psi'.  \eeq This defines the
vector operator $\L$, given component-wise by $ L_{i}
=\varepsilon_{ijk}p_k p_l {\partial }^2/{\partial x_{j}}{\partial
  x_{l}}$. Equation \eqref{AppEq:linconcstokes} then becomes
\begin{equation}
  \label{Appeq:lin1}
  \nabla^{4}\bPhi=\tilde{\alpha}\int_S dS_{p}\ \L \Psi'.
\end{equation}

Now we turn to the evolution equation \eqref{Appeq:linSmol}. The
contracted term $ \p \p : \nabla \u$ can be rewritten as $\p \p :
\nabla \u=-\L \cdot \bPhi$ (this uses that $\p \p: \E = \p \p: \nabla
\u$, and that $\varepsilon_{ijk}=-\varepsilon_{kji}$, followed by an
index relabeling). Then, Eq.~\eqref{Appeq:linSmol} can be rewritten as
\begin{equation}
\frac{\partial\Psi'}{\partial  t}=-\frac{3}{4 \pi} \L \cdot
\bPhi  + \frac{3 \zeta}{2 \pi} \p\p:\D' + d_T \Delta \Psi' +
d_R \nabla^2_{p}  \Psi'.
\label{Appeq:lin2a}
\end{equation}
Now we take a time derivative of Eq.~\eqref{Appeq:lin1}
(i.e. $\nabla^{4}\bPhi_t=\tilde{\alpha}\int_S dS_{p}\ \L \Psi'_t$), and
use Eqs.~\eqref{Appeq:lin2a} and \eqref{Appeq:lin1} to find
\begin{equation}
  \frac{\partial\nabla^4 \bPhi}{\partial t}=
 - \frac{3 \tilde \alpha}{4 \pi}  \int_S dS_{p}\ \L (\L \cdot \bPhi
  )+  \frac{3 \zeta \tilde \alpha}{2 \pi} \int_S dS_{p}\ \L (\p\p:\D') +
  d_T \nabla^6 \bPhi  +  {\tilde \alpha d_R}
  \int_S dS_{p}\ \L \  \nabla^2_{p}  \Psi' .
\label{Appeq:phit}
\end{equation}
The terms on the right-hand-side of Eq.~\eqref{Appeq:phit} can be calculated as
\begin{align}
  &\int_S dS_{p}\ \L  (\L \cdot \bPhi ) =-\curl \left( \nabla \cdot \int_S dS_{p}\ \left(\p\p \left( \nabla \u : \p\p \right)\right) \right) = \frac{4\pi }{15}\nabla ^{4}\bPhi, \\
  &\int_S dS_{p}\ \L\left(  \p\p^{T}:\D'\right)  = \frac{8\pi}{15} \int_S dS_{p}\ \L \Psi', \\
  &\int_S dS_{p}\ \L \  \nabla^2_{p}  \Psi' = - 6  \int_S dS_{p}\ \L  \Psi'.
\end{align}
Finally, by defining $\g = \nabla^4 \bPhi$ and using the definition for
${\tilde \alpha}$ the linearized dynamics reduces to
\begin{equation}
  \label{Appeq:b}
  \frac{\partial \g}{\partial  t} =  \left(-C_1(\beta)\alpha
  +C_2(\beta) \zeta -6d_R\right) \g + d_T  \Delta \g,
\end{equation}
where $C_1(z)=(3/5)/(1+z/15))$ and $C_2(z)=(4/5)+(6z/25)/(1+z/15)$.
Thus, growth or decay rates of plane-wave solutions are given by
\begin{equation}
\sigma=C_1(\beta)\alpha+C_2(\beta) \zeta -6d_R
-d_T k^2~.
\label{Appgrowth rates}
\end{equation}

\section{The dynamics of Model I in a circular domain}
\label{AppendixC}
\begin{figure}
\begin{center}
  \includegraphics[width=1.0 \textwidth]{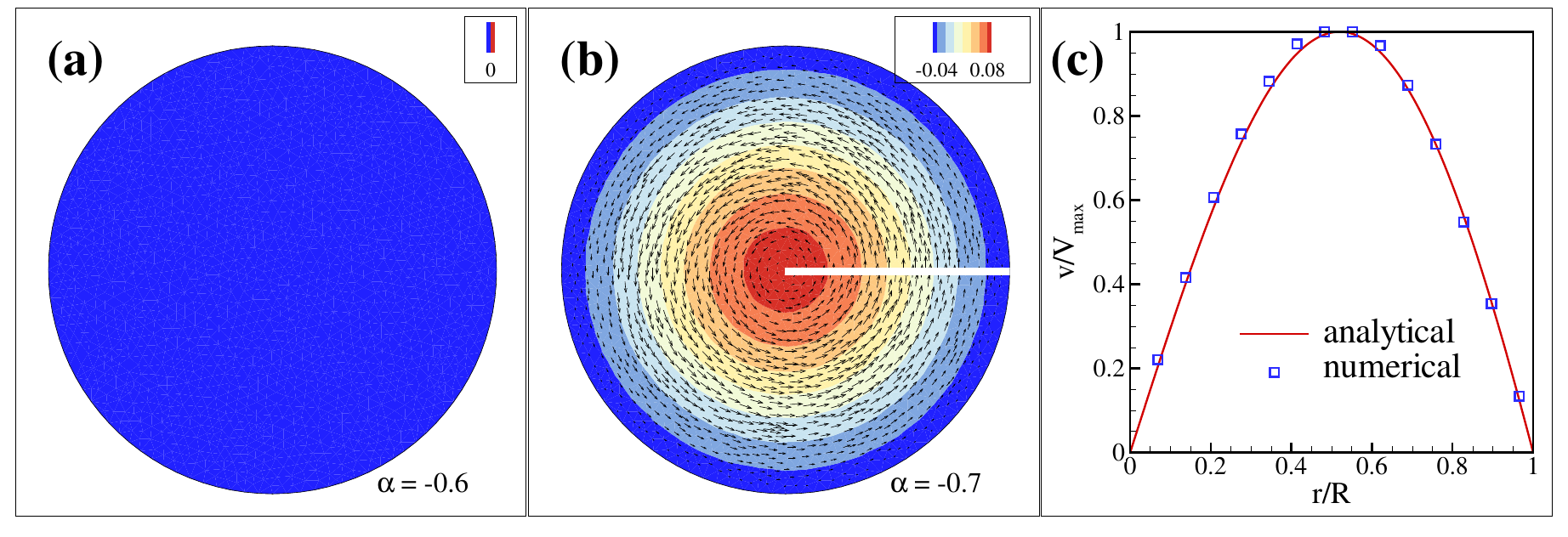}
\end{center}
\caption{Panels (a) and (b) show the fluid velocity vector field $\u$
  and its scalar vorticity $\omega$ (color field) for two Extensor
  suspensions confined to a circular chamber of radius $R=2$ using
  Model I. Panel (a) uses $\alpha = -0.6$, while (b) uses $\alpha =
  -0.7$, which spans the critical value $\alpha \approx -0.63$
  calculated by Woodhouse and Goldstein \cite{woodhouse2012}. Panel
  (a) shows the result of relaxation to a quiescent flow, while panel
  (b) shows the emergence of an auto-circulating flow.  Panel (c)
  shows a comparison of the analytical calculation for the azimuthal
  velocity in (b), with the numerically determined value taken along
  the white line in (b). }
  \label{fig:dilutecheck}
\end{figure}

\begin{figure}
\begin{center}
  \includegraphics[width=1.0 \textwidth]{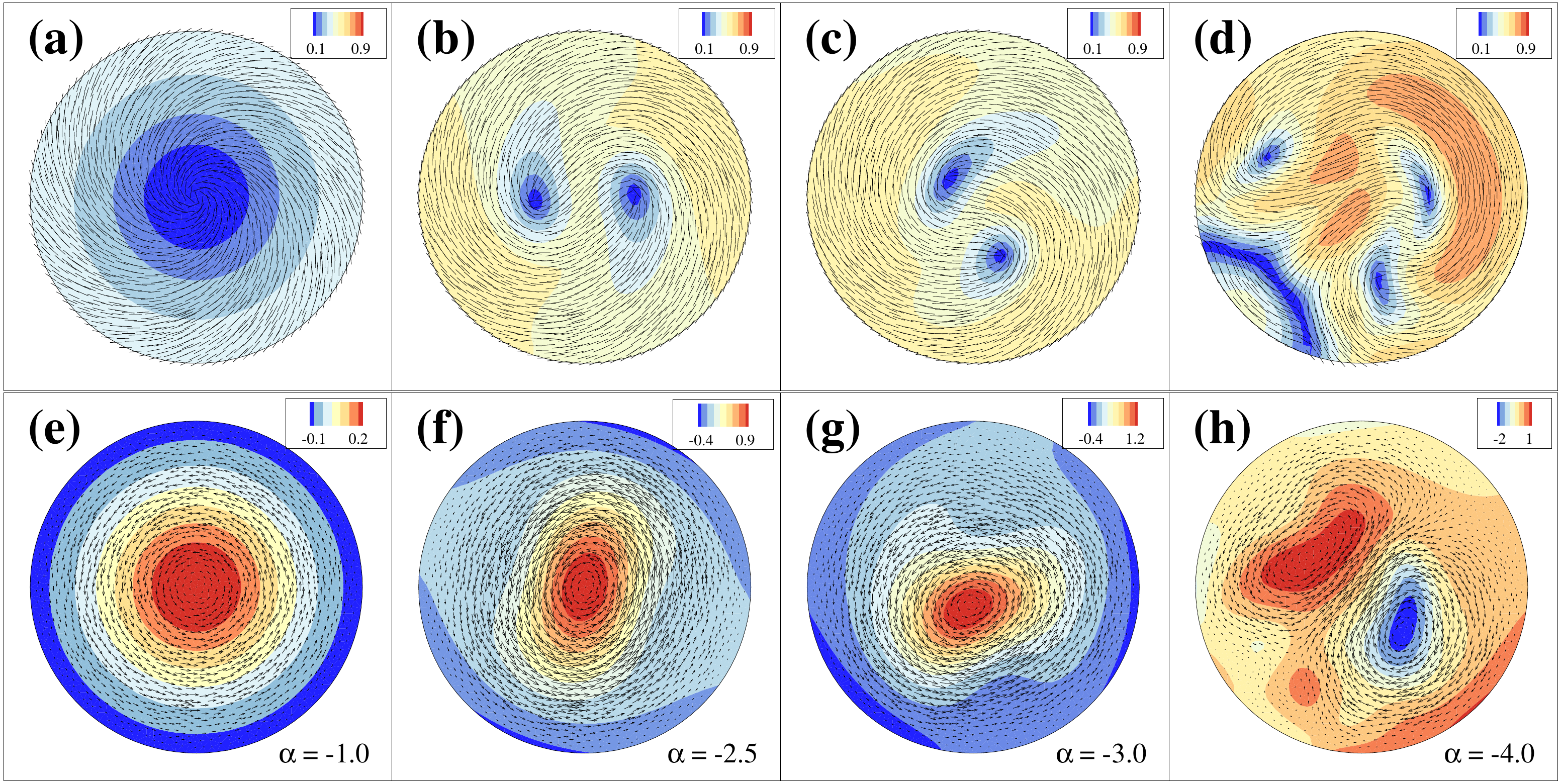}
\end{center}
\caption{Snapshots of the characteristic nematic and flow structures of Extensor
  suspensions of increasing activity(i.e., decreasing $\alpha$), when confined to a circular chamber of radius $R=2$. The upper row shows the nematic field $\m$ overlaying the scalar order
  parameter $s$ (color field), while the lower row shows the velocity
  vector field $\u$ overlaying its scalar vorticity (color
  field).}
  \label{fig:dilutedfct}
\end{figure}

\begin{figure}
\begin{center}
  \includegraphics[width=0.45 \textwidth]{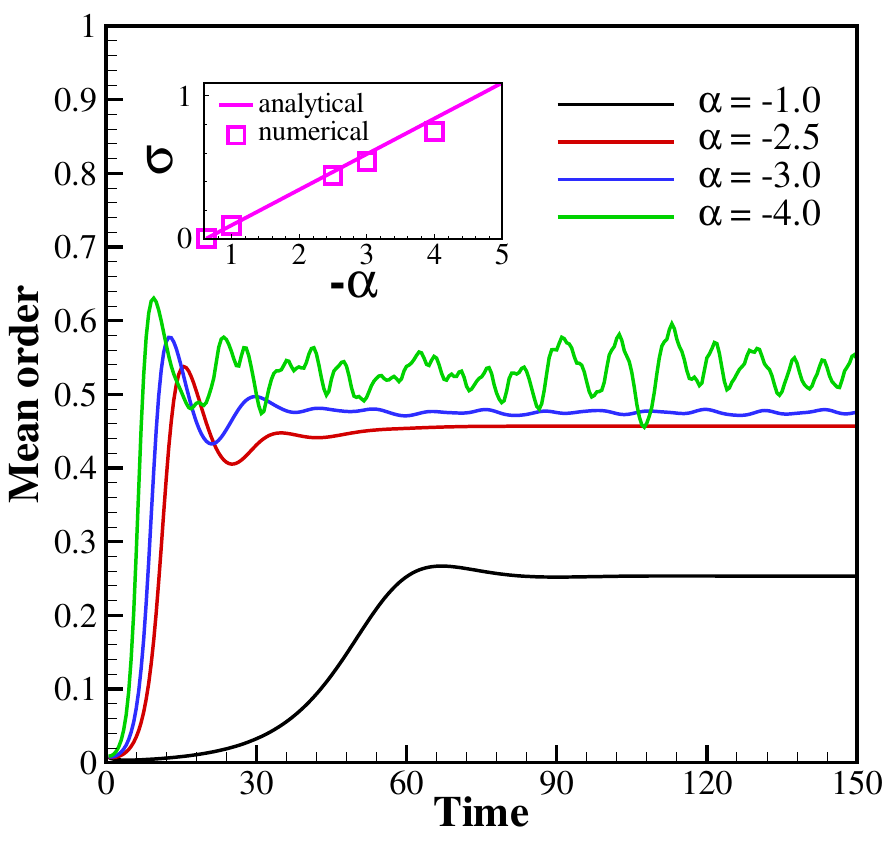}
\end{center}
\caption{Corresponding to the simulations shown in
  Fig.~\ref{fig:dilutedfct}, the evolution of the mean nematic order
  $S(t)$ at various values of $\alpha$. Again, the initial growth of $S$ is
  found to be exponential in time, and the inset shows a fit to the
  growth rate $\sigma$ (open squares) as a function of $-\alpha$. The solid
  line is the growth rate predicted by the analytical axisymmetric mode:
  $\sigma=-\frac{\alpha}{4}-4d_R-\left(\frac{\lambda_0}{R}\right)^2d_T$.}
  \label{fig:meanorder_dilute}
\end{figure}

We confine an Extensor suspension to a
circular chamber of radius $R$, starting the simulation near uniform
isotropy in the weakly aligned limit with $\zeta=\beta=0$, which
corresponds to neglecting steric interactions between filaments. This
limit corresponds to the model previously studied by Woodhouse and
Goldstein \cite{woodhouse2012}, who discovered a symmetry-breaking
bifurcation at which the stationary state transitions to a stable
circulating state. For higher activity, another bifurcation produces
an oscillatory state with a pair of mobile defects.

We expected that our Bingham closure would reproduce the Woodhouse and
Goldstein results that used the Hinch and Leal closure
\cite{hinch1976} when neglecting steric interactions. Indeed, we
recover an identical system of linearized equations for perturbation
about an isotropic state. Numerically, we capture the bifurcation to a
circulatory state (Fig.~\ref{fig:dilutecheck}), and
the growth rates agree quantitatively with linear stability analysis
(Fig.~\ref{fig:meanorder_dilute}). The shape of the flow profile
matches that determined analytically \cite{woodhouse2012}
(Fig.~\ref{fig:dilutecheck}).

For increasing activity (i.e., decreasing $\alpha$), in Fig.~\ref{fig:dilutedfct} we reproduce the oscillatory two-defect state
found by Woodhouse and Goldstein \cite{woodhouse2012}. For
sufficiently high activity, this oscillatory state is unstable to the
production of cracks and defects in the nematic field, and multiple
vortices.

\bibliography{zoterolibrary}{}
\bibliographystyle{unsrt}

\end{document}